\documentclass[aps,pre,reprint,showpacs,floatfix]{revtex4-1}
\usepackage{graphicx}
	
\usepackage{graphics}
\usepackage{epstopdf}
\usepackage{epsfig}
\usepackage{natbib}

\begin{document}

\title{Microtubule catastrophe from protofilament dynamics}

\author{V. Jemseena}
\email{jemseena@physics.iitm.ac.in}
\author{Manoj Gopalakrishnan}
\email{manojgopal@iitm.ac.in (corresponding author)}

\affiliation{Department of Physics, Indian
  Institute of Technology Madras, Chennai 600036,
  India}

\date{\today}

\begin{abstract}
The disappearance of the guanosine triphosphate (GTP)-tubulin cap is widely believed to be the forerunner event for the growth-shrinkage transition (`catastrophe') in microtubule filaments in eukaryotic cells. We study a discrete version of a stochastic model of the GTP cap dynamics, originally proposed by Flyvbjerg, Holy and Leibler (Flyvbjerg, Holy and Leibler, Phys. Rev. Lett. {\bf 73}, 2372, 1994). Our model includes both spontaneous and vectorial hydrolysis, as well as dissociation of a non-hydrolyzed dimer from the filament after incorporation. In the first part of the paper, we apply this model to a single protofilament of a microtubule.  A catastrophe transition is defined for each protofilament, similar to the earlier one-dimensional models, the frequency of occurrence of which is then calculated under various conditions, but without explicit assumption of steady state conditions. Using a perturbative approach, we show that the leading asymptotic behavior of the prot
 ofilament catastrophe  in the limit of large growth velocities is remarkably similar across different models. In the second part of the paper, we extend our analysis to the entire filament by making a conjecture that a minimum number of such transitions are required to occur for the onset of microtubule catastrophe. The frequency of microtubule catastrophe is then determined using numerical simulations, and compared with analytical/semi-analytical estimates made under steady state/quasi-steady state assumptions respectively for the protofilament dynamics. A few relevant experimental results are analyzed in detail, and compared with predictions from the model. Our results indicate that loss of GTP cap in 2-3 protofilaments is necessary to trigger catastrophe in a microtubule. 
 \end{abstract}


\maketitle

\section{Introduction}

Microtubules exhibit an intrinsic property whereby they switch between states of growth and shrinkage constantly. In the growing state, $\alpha-\beta$ tubulin dimeric units with $\beta$ tubulin carrying rapidly hydrolyzable  GTP  are added to the tip, thereby increasing the microtubule length. Since, microtubule lattice with GTP -bound tubulin is more stable than the one with GDP (guanosine diphosphate)-bound tubulin, hydrolysis leaves the microtubule unstable and eventually causes depolymerization of polymer. In the depolymerizing state, the hydrolyzed dimeric units are lost from the tip and results in shrinkage of microtubule. Thus, a given microtubule in a population would appear to be in either growing state or shrinking state, with alternate transitions between these two states. A third state called ``pause" has been observed both {\it in vivo} and {\it in vitro}, where a microtubule neither grows or shrinks. 

The GTP cap theory successfully explained the stochastic nature of microtubule dynamics, according to which a growing filament is characterized by the presence of a cap of GTP tubulin at its tip. The filament will keep growing as long as this cap is intact, even if most of the tubulin in the interior is in GDP bound, hydrolyzed state. Upon the loss of this temporary GTP cap by spontaneous and irreversible hydrolysis, the GDP rich region becomes exposed and the polymer undergoes depolymerization. This transition from a growing state to shrinking state is known as {\it catastrophe}. The reverse could happen, when the cap reappears at the tip either by addition of GTP-bound dimers or by exposure of a GTP remnant within, which leads to a transition from shrinking state to growing state: this is known as {\it rescue}. This collective dynamic behavior of a microtubule, consisting of alternating catastrophe and rescue  events is known as {\it dynamic instability} \cite{desai,mitchis
 on}. A thorough understanding of dynamic instability, which is the highlight of microtubule dynamics, is a key to understand microtubule dependent functions in biological systems \cite{howardreview2012}.    

The GTP cap theory itself does not specify the structure of the cap itself. As the microtubule itself consists of  a number (usually 13) of protofilaments arranged in parallel, it is conceivable that some of the protofilaments will be GTP-tipped, while some will not be. It is a matter of debate as to how many protofilaments are required to be GTP-tipped so as to 
define a cap; all 13 or less? And even in the former case, the GTP region at the tip will have, in general, variable lengths across different protofilaments. How do we define the length of the cap, in such a situation? { \it In vitro} experimental studies performed using the slowly hydrolyzable GTP analogue GMPCPP by Caplow  et al. have estimated the size of the GTP cap necessary to stabilize the polymer and it has been shown that a single layer of GTP  tubulin at the tip is sufficient \cite{caplow}. Later on, the experimental observations at nanoscale resolution by Schek  et al. suggest that the cap consists of multilayer of GTP tubulin with an exponentially distributed multilayer GTP cap \cite{schek}. But { \it in vivo}, the exact definition of cap that explains  whether it requires capping of all the thirteen protofilaments or not remains uncertain. Therefore, a more quantitatively precise characterization of the GTP cap theory and its implications for microtubule dynamics
  requires development of physical and mathematical models based on it, and comparison of the predictions of such models with experimental observations. 

A large variety of approaches have been adopted to tackle the problem of microtubule dynamics based on the GTP cap theory \cite{hill,FHL,margolin,antal,ranjith,vanburen,molodtsov,piette,nedelec,alber}, which differ essentially in the level of molecular details included in the respective models. As recently discussed by Margolin et al. \cite{alber}, the results from all models, by parameter tuning, appear to agree with available experimental observations irrespective of the details of microtubule structure included, and irrespective of the differences in mathematical expressions for (primarily) the frequency of catastrophe or related quantities. We feel that a comparative study of at least some of the models, from a common starting point, is desirable, and this is one of the objectives of this paper. We chose the complete stochastic model, first proposed by Flyvbjerg et al. \cite{FHL} as this starting point, as this model  appears to be fairly successful in reproducing many ob
 served features of microtubule catastrophe, from a purely kinetic point of view. However, almost all the mathematical models, including the original FHL model, are effective one-dimensional models where the microtubule is approximated to a linear polymer. They also differ in some important details: whereas some models neglect rescue altogether, some others ignore vectorial hydrolysis; dissociation of tubulin dimers from the filament in the growing state is included in some models, but not in others. As a result, the predictions are apparently different even across models which share a lot of similarities in their basic assumptions; therefore, we feel that a fresh study of the stochastic model is not untimely.

Flyvbjerg, Holy and Leibler \cite{FHL} (henceforth referred to as FHL) formulated an effective stochastic model of microtubule catastrophes, based on the GTP cap theory. In this model, GTP bound tubulin polymerizes to form a one dimensional filament, and may undergo hydrolysis anywhere in the polymer. Two types of hydrolysis processes were considered, spontaneous and induced (vectorial). While the former occurs in a GTP-bound tubulin independent of the nucleotide state of the neighbors, in the latter case, a GDP tubulin was assumed to enhance the rate of hydrolysis in a GTP-neighbor. The model made a number of predictions which compared favorably with experimental observations. In particular, the model predicted that the catastrophe frequency is a decreasing function of the growth velocity $v_g$, and asymptotically approaches a small, constant value as $v_g\to\infty$. The latter feature was a surprising, even counter-intuitive prediction, which we will show later to be an art
 ifact of the continuum theory used by FHL. The FHL model inspired a number of later studies, which attempted to go beyond its limitations, notably the papers by Antal et al. \cite{antal} and Padinhateeri et al. \cite{ranjith}. While the latter focused on frequencies of catastrophe and rescue, the former also studied statistical characteristics such as the length distributions of the 
GTP cap as well as the interior islands of T-mers and D-mers. Similar studies have also been carried out in the context of actin filaments  \cite{Li,ranjith2010}. We shall also give a comparison of our results with the studies by Antal et al. \cite{antal} and Padinhateeri et al. \cite{ranjith} in one of the later sections. 

The work on this paper was started with the objective of extending the FHL model in such a way so as to base it on the dynamics of individual protofilaments, and therefore to formulate it as a complete three-dimensional model of catastrophe. In doing so, we are also freed from the necessity of taking a spatially continuum approach in describing the hydrolysis process: as each protofilament is a single polymer with monomer molecules arranged linearly, a one-dimensional discrete formalism is easily implemented for the dynamics of each. In this way, local catastrophe events are defined for each protofilament (corresponding to the loss of the GTP-tip for each) the frequency of which is determined precisely under steady state conditions. The global/microtubule catastrophe is defined as an event that pertains to the entire microtubule, and whose onset was defined phenomenologically in terms of the number of individual protofilaments that underwent local catastrophes. This approach 
 gives us enough flexibility with the definition of global catastrophes, so that its dynamics (the time scale of which is much larger than the local catastrophes in each protofilament) as well as the steady state value could be studied in detail, and compared with experiments. Indeed, several experiments have highlighted the age dependence of microtubule catastrophe frequency; in general, following nucleation or rescue, catastrophe frequency is found to increase with time and saturate at a steady state value. By comparing results from our numerical simulations with recent experimental observations, we predict that the microtubule catastrophe requires at least 2-3 protofilaments to have lost their GTP tips.
\section {Model and definitions}

\begin{figure}[h]
 \centering
 \includegraphics[scale=.3,keepaspectratio=true]{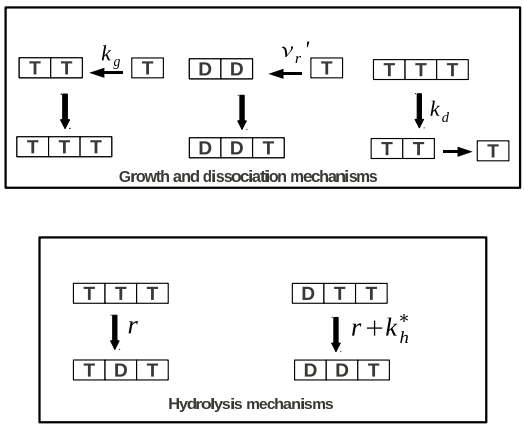}
\caption{ The figure is an illustration of the stochastic events in the dynamics of a single protofilament.}
 \label{fig:fig1+}
\end{figure}

A single protofilament in our model is a linear polymer of tubulin molecules, starting with one GTP-bound tubulin (symbolically denoted `T', and henceforth referred to as a T-mer). Further incoming T-mers are added to the protofilament at a rate $k_g$ if it is T-tipped, and at a rate $\nu_r^{\prime}$ if it is D-tipped. The rate of spontaneous hydrolysis is denoted $r$ and the rate of vectorial/induced hydrolysis is denoted $k_h^*$.  This means that in a ....TTT... configuration, the middle T becomes  D at a rate $r$, whereas in a ....DTT.. or ...DTD... configuration, the middle T becomes D at a rate $r+k_h^*$.  We may further allow for the possibility that a T-mer may detach from the tip at a rate $k_d$.  All these possible transitions with respective rates are schematically shown in Fig 1.  When the last T with a D neighbour is also lost by hydrolysis or detachment from the tip, we define the protofilament to undergo a `local catastrophe'. The rate at which the transition fr
 om growing phase to shrinking phase occurs is denoted by $\nu_c^{\prime}$, which we refer to as the frequency of `protofilament catastrophe'.  Analogously, $\nu_r^{\prime}$ may be defined as the `protofilament rescue' \cite{foot}, which we regard as a time independent constant throughout this work. Of these parameters, the rates $k_g, r,k_h^*, k_d$ and $\nu_r^{\prime}$ are independent parameters which may be regarded as constants, while $\nu_c^{\prime}$ needs to be computed in terms of these, and is, in general, a time dependent quantity. 

The complete microtubule may be imagined as a set of 13 such protofilaments arranged side by side (Fig 2). Our approach in this paper is essentially kinetic in nature, and we do not propose to undertake a detailed treatment of the energetics in the problem, which has been carried out by several authors \cite{vanburen,molodtsov,piette,alber}. For this reason, 
we do not consider explicitly the energy of interaction between protofilaments or the bending energy of individual protofilaments. Hence the cylindrical structure of the entire microtubule filament is irrelevant in our model, where the microtubule `lattice' has a flat geometry, similar to the model studied in Ref. \cite{nedelec}. 

In order to define a catastrophe event for the entire filament, we adopt the following phenomenological definition. We conjecture that when a certain minimum number $n^*$ of individual protofilaments have undergone their individual catastrophes, the entire filament becomes energetically unstable and enters the shrinking phase. Therefore, this `first passage' event where $n^*-1$ protofilaments have already undergone catastrophes by time $T$ (and not rescued until the time instant $T$) while the $n^*$'th one undergoes catastrophe at time $T$ defines a catastrophe event for the filament at time $T$, and denote it by $\nu_c$. One of our objectives of this work is to estimate the number of protofilaments that needs to undergo local catastrophe to produce a catastrophe of the filament. 

\begin{figure}[h]
 \centering
 \includegraphics[scale=.23,keepaspectratio=true]{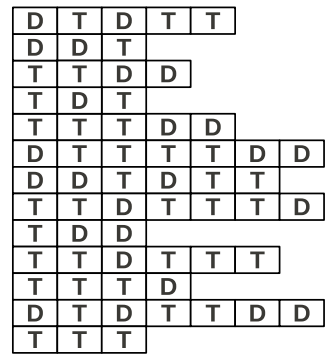}
\caption{The `flat' microtubule model studied in this paper, showing 13 protofilaments arranged side by side.}
 \label{fig:fig1}
\end{figure}

\section{Dynamical equations}

\subsection{The Master Equation}

Let $P_n(t)$ the probability that a protofilament has  a GTP cap consisting of $n$ T-mers at time $t$. $P_n(t)$ evolves in time either by addition or loss of monomers. Monomers are added to the tip of the cap at a rate $k_g$. The size of the cap becomes smaller by hydrolysis of GTP to GDP at the inner boundary at  rate $k_h^*+r$, or anywhere within the cap at rate $r$ or by detachment of T-mer from the tip at rate $k_d$ . Since the effect of detachment process is the same as that of vectorial hydrolysis, we may simply effect a replacement $k_h^*\to k_h\equiv k_h^*+k_d$ in the equations to take this into account.

We assume next that a protofilament in which the cap length has shrunk to zero would undergo rescue at a rate $\nu_r^{\prime}$, and that the rescued protofilaments continue to grow by successive addition of monomers, until it encounters the next catastrophe event. Hence, to begin with, we consider  an ensemble of protofilaments such that, at $t=0$, a protofilament possess a cap with $N$ subunits. The cap dynamics can be summarized into the following master equation: 

\begin{eqnarray}
\frac{dP_n(t)}{dt} =k_h[P_{n+1}(t)-P_{n}(t)]+ k_g[P_{n-1}(t)-P_{n}(t)]\nonumber\\
-r n P_{n}(t)+r\sum_{m=n+1}^{\infty} P_{m}(t) ~~~~~~~n\geq2 \nonumber \\
\frac{dP_1(t)}{dt} =k_h[P_{2}(t)-P_{1}(t)]- k_g P_{1}(t)+\nu_r^{\prime} P_0(t) \nonumber \\
-r P_{1}(t)+r\sum_{m=2}^{\infty} P_{m}(t) \nonumber\\
\frac{dP_0(t)}{dt} =k_hP_1(t)-\nu_r^{\prime} P_0(t)+r\sum_{m=1}^{\infty} P_{m}(t). ~~~
\label{eq:eq1}
\end{eqnarray}

The distribution $P_n(t)$ is normalized: $\sum_{n=0}^{\infty}P_n(t)=1$ at all times $t$, consistent with the above equation. The third and fourth term containing $r$ in the equation for $P_n(t)$ ($n\geq2$) can be explained respectively as follows. Protofilaments with $n$ T-mers at the tip can switch to a state with tip consisting of less than $n$ T-mers caused by the spontaneous hydrolysis, which cuts the GTP cap in to two regions of T-mers separated by a D-mer. By the reverse process, a protofilament of cap length  $n$ can be generated from one with a cap of length larger than $n$. In this way  the spontaneous hydrolysis accelerates the stochastic switching between growing and shrinking states.      

We now define a catastrophe event for a protofilament, and derive an expression for the frequency of occurrence of the same. In conformity with the definition given in the last section, we define protofilament catastrophe frequency as 

\begin{equation}
 \nu_{c}^{\prime}(t)=\frac{(r+k_h)P_1(t)}{1-P_{0}(t)}\leq r+k_h,
 \label{eq:eq2}
\end{equation}
where the numerator gives the fraction of protofilaments undergoing catastrophe in the interval $[t:t+dt]$ while the denominator gives the fraction that is in growing state (i.e., with a GTP-tip of non-zero length) at time $t$. The upper bound in Eq.\ref{eq:eq2} follows from normalization.

\subsection{Perturbation Theory}

The occurrence of the non-local terms in the sums in Eq.\ref{eq:eq1} means that finding a general solution to this equation is likely to be cumbersome. However, we note that when $r=0$, the set of equations in Eq.\ref{eq:eq1} describes a (discrete) random walk in one dimension with a boundary condition at $n=0$, which can be exactly solved. 
 It is therefore logical to use a perturbative method, where the distribution $P_n(t)$ is expanded in powers of $r$. Also, it can be shown by scaling arguments (see subsection E later) that in the asymptotic $k_g\to \infty$ regime, the leading term in the expression for $\nu_c^{\prime}$ is indeed the first order perturbation term in $r$. Therefore, we now start with the expansion 

\begin{equation}
P_n(t)=P_n^{(0)}(t)+rP_n^{(1)}(t)+r^2P_n^{(2)}(t)+\dots,
\label{eq:eq3}
\end{equation}
$P_n^{(0)}(t)$, which describes dynamics without spontaneous hydrolysis, satisfies the following equations:

\begin{eqnarray}
\frac{dP_n^{(0)}(t)}{dt} =k_h[P_{n+1}^{(0)}(t)-P_{n}^{(0)}(t)]\nonumber\\ 
+ k_g[P_{n-1}^{(0)}(t) -  P_{n}^{(0)}(t)] ~~ n\geq2\nonumber\\
\frac{dP_1^{(0)}(t)}{dt} =k_h[P_{2}^{(0)}(t)-P_{1}^{(0)}(t)]- k_g P_{1}^{(0)}(t)+\nu_r^{\prime} P_0^{(0)}(t)\nonumber\\
\frac{dP_0^{(0)}(t)}{dt} =k_hP_{1}^{(0)}(t)-\nu_r^{\prime} P_0^{(0)}(t),~~~~~
\label{eq:eq4}
\end{eqnarray}
while $P_n^{(1)}(t)$ satisfies the equations

\begin{eqnarray}
 \frac{dP_n^{(1)}(t)}{dt} =k_h[P_{n+1}^{(1)}(t)-P_{n}^{(1)}(t)]+ k_g[P_{n-1}^{(1)}(t)\nonumber\\-P_{n}^{(1)}(t)]
-n P_{n}^{(0)}(t)+\sum_{m=n+1}^{\infty} P_{m}^{(0)}(t) , n\geq2 \nonumber\\
\frac{dP_1^{(1)}(t)}{dt} =k_h[P_{2}^{(1)}(t)-P_{1}^{(1)}(t)]- k_g P_{1}^{(1)}(t)\nonumber \\+\nu_r^{\prime} P_0^{(1)}(t)-P_{1}^{(0)}(t)+\sum_{m=2}^{\infty} P_{m}^{(0)}(t)\nonumber\\
 \frac{dP_0^{(1)}(t)}{dt} =k_hP_{1}^{(1)}(t)-\nu_r^{\prime} P_0^{(1)}(t)+\sum_{m=1}^{\infty} P_{m}^{(0)}(t). 
\label{eq:eq5}
\end{eqnarray} 

Further, normalization needs to be satisfied for all $r$, which requires that $\sum_{n=0}^{\infty}P_n^{(0)}(t)=1$ while $\sum_{n=0}^{\infty}P_n^{(k)}(t)=0$ for $k\geq 1$. In order to solve for $P_n(t)$, we use the generating function, defined as 

\begin{equation}
\phi(z,t)=\sum_{n=0}^{\infty} z^n P_n(t),
\label{eq:eq6}
\end{equation}
with the  inversion formula 

\begin{equation}
 P_n(t)=\frac{1}{2\pi i}\oint \frac{\phi (z,t)dz}{z^{n+1}}.
 \label{eq:eq7}
\end{equation}

In the above expression, the contour is taken as a circle of infinitesimal radius centered at the origin $z=0$. The generating function itself has  the perturbation theory expansion $\phi(z,t)=\phi^{(0)}(z,t)+r\phi^{(1)}(z,t)+.......$. Naturally, the inversion formula in Eq.\ref{eq:eq7} also applies to each order in $r$:

\begin{equation}
 P_n^{(k)}(t)=\frac{1}{2\pi i}\oint \frac{\phi^{(k)} (z,t)dz}{z^{n+1}}~~~~~k\geq 0.
\label{eq:eq8}
\end{equation}

Using the power-series expansion in Eq.\ref{eq:eq3}, the catastrophe frequency in Eq.\ref{eq:eq2} may be expanded in the form

\begin{eqnarray}
\nu_c^{\prime}(t)=\frac{k_hP_1^{(0)}(t)}{1-P_0^{(0)}(t)}+r\bigg\{\frac{P_1^{(0)}(t)}{1-P_0^{(0)}(t)}+\frac{ k_hP_1^{(0)}(t) P_0^{(1)}(t)}{(1-P_0^{(0)}(t))^2}\nonumber\\+\frac{k_hP_1^{(1)}(t)}{1-P_0^{(0)}(t)}
\bigg\}+O(r^2).~~~~~
\label{eq:eq9}
\end{eqnarray}

In order to calculate the protofilament catastrophe using Eq.\ref{eq:eq9}, we first solve Eq.4-5 by defining the Laplace transforms ${\tilde P}^{(k)}_n(s)=\int_{0}^{\infty}P_n^{(k)}(t)e^{-st}dt$.  The general expressions for ${\tilde P}_{n}^{(k)}(s)$ with $n,k=0,1$ (relevant to Eq.\ref{eq:eq9}) as well the relations between them are given in the Appendix.

\section{Results I: Protofilament catastrophe}

We will first calculate the protofilament catastrophes in the steady state limit under different regimes, which are conveniently classified as below:

When the protofilament rescue rate $\nu_r^{\prime}>0$, the system reaches a complete steady state, where all the probabilities $P_n$ become independent of time in the long-time 
limit, and so does the catastrophe. Here, we study the cases $k_g>k_h$ and $k_g<k_h$ separately (a demarcation warranted by perturbation theory, but seemingly artificial, since  numerical solution of the master equation shows that the catastrophe frequency varies continuously across  $k_g=k_h$, see Fig.\ref{fig:fig2} later). When $\nu_r^{\prime}=0$, on the other hand, the $n=0$ state becomes absorbing and hence none of the probabilities $P_n (n\geq 1)$ has a non-zero steady state value.  However, even in this case, the catastrophe frequency is found to have a well-defined non-zero steady state value, which is different from the previous case. These cases are treated in subsections A and B respectively. Finally, it is seen that the special case $k_h=0$ can be solved exactly for both $\nu_r^{\prime}>0$ and $\nu_r^{\prime}=0$, see subsection D.

In order to find the steady state value of $\nu_c^{\prime}(t)$, the steady state values of all dynamical quantities appearing in Eq.9 are found in the $t\to\infty$ limit. We denote the steady state values of $P_n(t)$ and $\nu_c^{\prime}(t)$ by the same symbols, but without the $t$-dependence. Using Laplace transforms, these limits may be defined as $P_n^{(k)}=\lim_{s\to 0}s{\tilde P}_{n}^{(k)}(s)$. If the limit turns out to be zero, the steady state value is zero (i.e., the corresponding dynamical quantity vanishes as $t\to\infty$ and a more careful treatment will be needed to understand the $t\to\infty$ behavior, as is required when $\nu_r^{\prime}=0$). 

Given that calculations involved are somewhat lengthy, we only give a summary of our final results in the main text, while the mathematical details are presented in the supplemental material to the paper \cite{supp}.

\subsection{Non-zero rescue: $\nu_r^{\prime}>0, k_h>0$}

\subsubsection{$k_h>0, k_g>k_h$}

The steady state protofilament catastrophe frequency  takes the form

\begin{equation}
 \nu_{c}^{\prime}=\frac {rk_h(2k_g-k_h)}{(k_g-k_h)^2}+O(r^2).
 \label{eq:eq13}
\end{equation}

Interestingly, we note that the expression in Eq.\ref{eq:eq13} differs from the corresponding expression for the microtubule catastrophe frequency in the one-dimensional effective continuum FHL model, which  (in our notation) is given by

 \begin{equation}
 \nu_c^{\mathrm{FHL}}=\frac {r^{\prime}\delta x(k_g+k_h)}{2(k_g-k_h)}+O({r^{\prime}}^2),
\label{eq:eq14}
 \end{equation}
 where $r^{\prime}$ is the spontaneous hydrolysis rate per unit length. A brief derivation of the above result, under perturbation theory, is given in the supplemental material \cite{supp}.  Asymptotically, while the continuum model predicts that $\nu_c^{\prime}\to r^{\prime}\delta x/2$ as $k_g\to\infty$, the discrete model predicts that $\nu_c^{\prime}$ vanishes in this limit as $\sim 2rk_h/k_g$. 

Both the continuum and discrete expressions diverge at $k_g=k_h$, however, this divergence is not real. A scaling argument (see later) shows that, as we approach the point $k_g=k_h$, the higher order terms in $r$ begin to be important, and are also possibly divergent as $k_g\to k_h$. If such terms have alternating signs, the complete function may well be convergent and well-behaved at the point $k_g=k_h$. Indeed, numerical solution confirm this argument. 

We next consider the case of small growth velocities.

\subsubsection{$k_h>0, k_g<k_h$}

In this case, detailed calculations show that the steady state catastrophe frequency takes the form 

\begin{eqnarray}
 \nu_{c}^{\prime}=k_h-k_g+r\bigg\{\frac{(k_h-k_g)}{k_h}+\nonumber\\
 \bigg(\frac {(k_h+\nu_r^{\prime}-k_g)}{\nu_r^{\prime}}\bigg)\bigg[\frac{k_g}{(k_h-k_g)}-\frac{k_g}{(k_h+\nu_r^{\prime}-k_g)}\nonumber \\+\frac{k_g^2}{(k_h-k_g)^2}+\frac{kg(\nu_r^{\prime}-k_g)}{(k_h+\nu_r^{\prime}-k_g)(k_h-k_g)}\bigg]\bigg\}+O(r^2).~~~\nonumber \\
 \label{eq:eq15}
\end{eqnarray}

Unlike the previous case, here, the catastrophe frequency has a non-vanishing zero'th order term, which decreases linearly with $k_g$ and vanishes at $k_g=k_h$.  The first order term, however, again diverges at $k_g=k_h$ (which should be interpreted with the same reservations as previously). It is interesting to note that $\nu_c^{\prime}$ is dependent on the protofilament rescue frequency, and as $\nu_r^{\prime}\to 0$, the following limiting value is reached: 

\begin{equation}
\lim_{\nu_r^{\prime}\to 0}\nu_c^{\prime}=(k_h-k_g)+r\frac{(k_h^3-k_g^3-k_gk_h^2+2k_hk_g^2)}{k_h(k_h-k_g)^2}+O(r^2).~
\label{eq:eq16}
\end{equation}

 As $k_g\to 0$, $\nu_c^{\prime}\to r+k_h$, which is its maximum value (see Eq.\ref{eq:eq2}, and also Eq.\ref{eq:eq23}, later).

\subsection{ Zero rescue: $\nu_r^{\prime}=0,k_h>0$ }

Here, we consider the situation where the protofilament rescue rate is strictly zero. This `non-steady state' situation is the case studied in many theoretical models, including \cite{FHL}.

\subsubsection{$k_h>0,k_g>k_h$}

The case of zero rescue needs to be treated with caution, as, strictly speaking, the only steady state possible is $P_0=1$ and $P_n=0$ for $n\geq 1$. However, careful calculations using the perturbative approach shows that the catastrophe frequency does reach a steady state, and the expression turns out to be \cite{supp} 

\begin{equation}
 \nu_c^{\prime} =\frac{rk_h}{(k_g-k_h)}+{\cal O}(r^2).
 \label{eq:eq17}
\end{equation}

Note that Eq.\ref{eq:eq17} is different from the expression in Eq.\ref{eq:eq13}, and its asymptotic form is $\nu_c^{\prime}\sim rk_h/k_g$ as $k_g\to\infty$. The singularity at $k_g=k_h$ is again an artifact of the perturbation theory. 

The general solution for $k_g<k_h$ turned out to require  a lengthy calculation, and hence was not pursued; rather, we found it instructive to look at the extreme case of vanishing growth rate, i.e., $k_g=0$.

\subsection{ Zero growth: $k_g=0$}

When $k_g=0$, if the initial condition is $P_n(0)=\delta_{nN}$ for $n\geq 0$ and $N\geq 1$, the condition of zero growth then guarantees that $P_n(t)=0$ for $n>N$ at all times $t$. We may further assume that the steady state $\nu_c^{\prime}$ is independent of the initial value $N$; therefore, it may be obtained by solving Eq.\ref{eq:eq1} exactly with a small value of $N$; here we chose $N=3$. The exact time-dependent solutions for the relevant probabilities in this case are given as

\begin{eqnarray}
 P_1(t)=\frac{(k_h^2+2r^2+2k_hr)}{2r^2} e^{-(k_h+r)t} \nonumber \\-\frac {(k_h+r)^2}{r^2}e^{-(k_h+2r)t}+\frac{k_h(k_h+2r)}{2r^2}e^{-(k_h+3r)t}\label{eq:eq18}\\
 P_0(t)=1-\frac{(k_h^2+2r^2+2k_hr)}{2r^2} e^{-(k_h+r)t} \nonumber \\+\frac {k_h(k_h+r)}{r^2}e^{-(k_h+2r)t}-\frac{k_h(k_h+2r)}{2r^2}e^{-(k_h+3r)t},
\label{eq:eq19}
\end{eqnarray}
which yields, using Eq.\ref{eq:eq2}, 

\begin{eqnarray}
 \nu_c^{\prime}=r+k_h~~~(\nu_r^{\prime}=0,k_g=0),
\label{eq:eq20}
\end{eqnarray}
exactly. In fact, this technique may be used for the $\nu_r^{\prime}>0$ case also, at the point $k_g=0$. In this case, we arrive at the following exact steady state expressions 
for $P_1$ and $ P_0$, with the same initial condition:

\begin{eqnarray}
 P_1=\frac{\nu_r^{\prime}}{k_h+\nu_r^{\prime}+r}
\label{eq:eq21}\\
 P_0=\frac{k_h+r}{k_h+\nu_r^{\prime}+r},
\label{eq:eq22}
\end{eqnarray}
 which then yield the exact result
 
\begin{eqnarray}
 \nu_c^{\prime}=r+k_h~~~~(\nu_r^{\prime}>0,k_g=0).
\label{eq:eq23}  
\end{eqnarray}

 Thus, the extremal $k_g=0$ value for $\nu_c^{\prime}$ is the same for $\nu_r^{\prime}=0$ and $\nu_r^{\prime}>0$, but their asymptotic behavior at large $k_g$ differ by a factor of 2. The same result given by Eq.\ref{eq:eq23} can be obtained for $N=2,4,5$ etc, but the calculations become lengthier as $N$ increases.

\subsection{No vectorial hydrolysis and monomer dissociation: $k_h=0$} 

Finally, we consider the case where the $k_h$ term is absent from Eq.\ref{eq:eq1}; this implies that there is no vectorial hydrolysis in the model, and no dissociation of T-mers from the protofilament prior to catastrophe. This case has been studied by several authors \cite{antal,piette,nedelec} in recent times (Note that in Ref. \cite{ranjith}, vectorial hydrolysis is deemed absent, but T-mer dissociation is included). In this situation, it is possible to obtain exact solutions to Eq.\ref{eq:eq1}, and hence $\nu_c^{\prime}$ can be computed for arbitrary $r$. As earlier, we consider the cases $\nu_r^{\prime}>0$ and $\nu_r^{\prime}=0$ separately. 

\subsubsection{Rescue present: $\nu_r^{\prime}>0$}

A steady state is possible in this case, as can be obviously verified by putting the time derivatives in Eq.\ref{eq:eq1} to zero. The steady state values of $P_0$ and $P_1$ 
then turn out to be 
 
\begin{equation}
P_0=\frac{r}{\nu_r^{\prime}+r}~~~;~~~P_1=\frac{2r\nu_r'}{(k_g+2r)(\nu_r'+r)}.
\label{eq:eq24}
\end{equation}

Upon substituting Eq.\ref{eq:eq24} into Eq.\ref{eq:eq2}, we find that 

\begin{equation}
 \nu_c^{\prime}=\frac{2r^2}{(k_g+2r)}.
 \label{eq:eq25}
\end{equation}

Unlike all other cases studied so far, the steady state catastrophe rate is $O(r^2)$ in the perturbation series, and vanishes as $2r^2/k_g$ as $k_g\to\infty$. One cannot fail to note the 
surprising similarity with the asymptotic decay of the expression in Eq.\ref{eq:eq13}; the expressions become identical when $k_h$ is replaced with $r$ in Eq.\ref{eq:eq13}.

 \subsubsection{Rescue absent: $\nu_r^{\prime}=0$}
 
In this case, as $t\to\infty$, $P_0(t)\to 1$ while all $P_m(t)\to 0$ for $m\geq 1$. Therefore, we need to consider the time evolution of the probabilities explicitly. With initial conditions $P_0(t=0)=P_1(t=0)=0$ and $P_2(t=0)=1$, the time-dependence of $P_0$ and $P_1$, as found from Eq.\ref{eq:eq1} are given by 

\begin{equation}
 P_0(t)=1-e^{-rt}~~~;~~~ P_1(t)=\frac{r}{(k_g+r)}[e^{-rt}-e^{-(k_g+2r)t}],
 \label{eq:eq26}
\end{equation}
which, after substitution in Eq.\ref{eq:eq2} gives 

\begin{equation}
 \nu_c^{\prime}= \frac{r^2}{(k_g+r)}
\label{eq:eq27}
\end{equation}
 as the long-time limit of the catastrophe frequency. As we noticed in the comparison between Eq.\ref{eq:eq25} and Eq.\ref{eq:eq13}, the asymptotic value $r^2/k_g$ is similar to that of Eq.\ref{eq:eq17}, when the replacement $k_h\to r$ is done. Although the steady state expressions for $\nu_c^{\prime}$ in Eq.\ref{eq:eq13} or Eq.\ref{eq:eq25} do not depend on $\nu_r^{\prime}$, they are, nevertheless, different from the corresponding $\nu_r^{\prime}=0$ expressions in Eq.\ref{eq:eq17} and Eq.\ref{eq:eq27}. To our knowledge, this difference has not been noted earlier. 
 
 \subsection{Asymptotic behavior: $k_g\gg r,k_h$}
 
 The results from perturbation theory discussed so far are significant in another way; scaling arguments show that the leading steady state term in the perturbation theory expansion of $\nu_c^{\prime}$ also gives the leading asymptotic behavior of $\nu_c^{\prime}$ in the limit $k_g\to \infty$. We first note that, in Eq.\ref{eq:eq1}, it is possible to define a dimensionless time by dividing the entire equation by $k_g$, whence $t\to T=k_gt$, while $R\to {\tilde R}\equiv R/k_g$, where the rate $R$ could be $\nu_r^{\prime}, k_h$ or $r$. We then expect that the catastrophe frequency has the scaling form
 
 \begin{equation}
 \nu_c^{\prime}=k_gf({\tilde r},{\tilde k_h},{\tilde \nu_r^{\prime}}),
 \label{eq:eqXX1}
 \end{equation}
 and according to our original assumption, we expect that the scaling function $f$ admits a power-series expansion of the form 
 
 \begin{equation}
 f({\tilde r},{\tilde k_h},{\tilde \nu_r^{\prime}})= f(0,{\tilde k_h},{\tilde \nu_r^{\prime}})+{\tilde r}\partial_{\tilde r}  f({\tilde r},{\tilde k_h},{\tilde \nu_r^{\prime}})|_{\tilde r=0}+..........
 \label{eq:eqXX2}
 \end{equation}

From Eq.\ref{eq:eq13} and Eq.\ref{eq:eq17} we observe that, when $k_g>k_h$, with $k_h>0$, the first term in the above equation vanishes. We also observe that the first derivative term is singular at ${\tilde k_h}=1$ but well-behaved otherwise (in particular, when ${\tilde k_h}\to 0$), and we may expect that this is true for the subsequent derivatives too (considering the upper bound in Eq.\ref{eq:eq2}). Therefore, in the limit $k_g\to\infty$, the expression in Eq.\ref{eq:eqXX1} takes the form

\begin{equation}
 \nu_c^{\prime}\sim r\lim_{{\tilde k_h},{\tilde \nu_r^{\prime}}\to 0}\bigg[\partial_{\tilde r}  f({\tilde r},{\tilde k_h},{\tilde \nu_r^{\prime}})|_{\tilde r=0}+O\bigg(\frac{r}{k_g}\bigg)\bigg],
 \label{eq:eqXX3}
 \end{equation}
  which means that the $O(r)$ term  is the leading asymptotic term. 
  
Now, what happens if $\nu_r^{\prime}=0$ or $k_h=0$ strictly? In the first case, the asymptotic structure of Eq.\ref{eq:eqXX3} appears to be retained, but with a different function $f$ in Eq.\ref{eq:eqXX2} (compare Eq.\ref{eq:eq13} and Eq.\ref{eq:eq17}). However, if $k_h=0$, the first derivative term vanishes, and the second derivative term becomes the leading term and hence $\nu_c^{\prime}=O(r^2)$ (Eq.\ref{eq:eq25} and Eq.\ref{eq:eq27}). 

The preceding analysis shows that some caution is required when experimental data is used to infer about the existence or non-existence of vectorial hydrolysis. The asymptotic behavior of $\nu_c^{\prime}$ in the large $k_g$ limit is similar in both cases, and cannot be used to distinguish between them. Further, a nonzero rate of dissociation of T-mers from the protofilament has an effect identical to that of vectorial hydrolysis. 

As a second important observation, we note that protofilament rescue events (here, the incorporation of a T-mer to a D-tipped protofilament) are important in determining the frequency of catastrophe. Our analysis has shown that, in all cases, $\nu_c^{\prime}$ is independent of the precise value of $\nu_r^{\prime}$, but depends on whether $\nu_r^{\prime}$ is strictly zero or not.

We now combine the results in Eq.\ref{eq:eq13}, Eq.\ref{eq:eq17}, Eq.\ref{eq:eq25}, Eq.\ref{eq:eq27} with the scaling argument in Eq.\ref{eq:eqXX3} to arrive at the following universal asymptotic form for the protofilament catastrophe in the limit $k_g\gg \max(r,k_h)$:

\begin{equation}
\nu_c^{\prime}\sim \frac{\Gamma}{k_g}~~~~~k_g\to\infty.
\label{eq:eqXX4}
\end{equation}

For easy reference, the values taken by the constant $\Gamma$ in different situations are summarized in Table \ref{tab:tab1}. 

\begin{table}
\begin{tabular}{|c|c|c|}
\hline
$\Gamma$ & $k_h=0$ & $k_h>0$ \\
\hline
$\nu_r^{\prime}=0$ & $r^2$ & $k_h r$\\
\hline
$\nu_r^{\prime}>0$ & $2r^2$ & $2k_h r$\\
\hline
\end{tabular}
\caption{The table lists the value taken by the parameter $\Gamma$ (Eq.\ref{eq:eqXX4}) under various conditions on $k_h$ and $\nu_r^{\prime}$.}
\label{tab:tab1}
\end{table}

\subsection{Numerical solution of the master equation}

\begin{figure}[h]
 \centering
 \includegraphics[scale=.32,keepaspectratio=true]{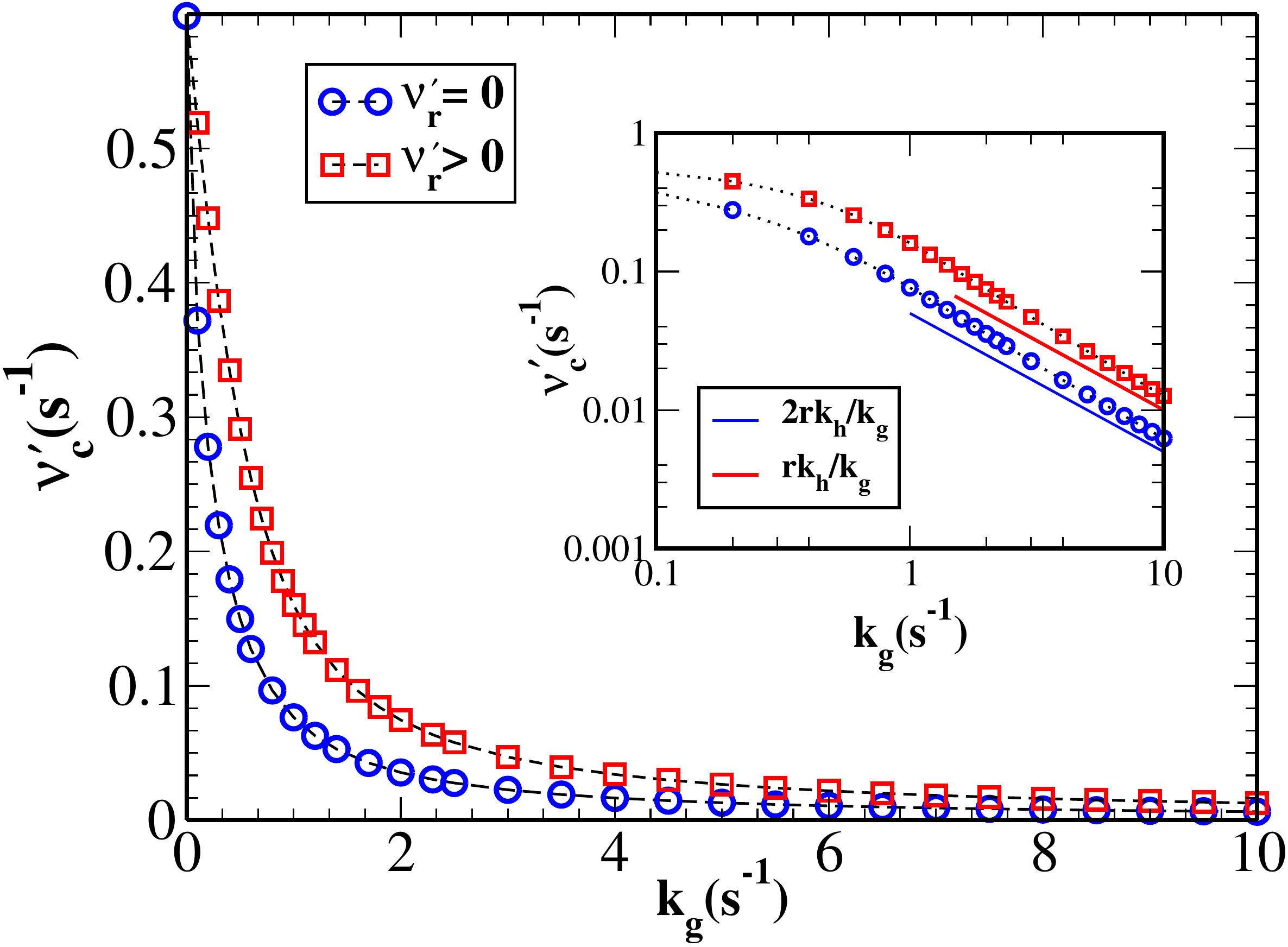}
 \caption{(color online) Protofilament steady state catastrophe $\nu_c^{\prime}$ versus $k_g$. The fixed parameters are $r=0.1\mathrm{s}^{-1}, k_h=0.5\mathrm{s}^{-1}$, 
 $\nu_r^{\prime}=0$ (blue circles) and $\nu_r^{\prime}=0.01\mathrm{s}^{-1}$ (red squares). The inset shows the same data in a double logarithmic scale. The lines (dashed blue line and dotted red line) are fits with slope $-1$, which confirms the 
 asymptotic behavior in both cases.}
 \label{fig:fig2}
\end{figure}

As the concluding step in our analysis, we  also solved Eq.\ref{eq:eq1} numerically using a simple Euler forward discretization scheme, with time interval $\delta t= 10^{-4}$s and the initial condition fixed to be $N=3$. The time-dependent catastrophe frequency  was computed using Eq.\ref{eq:eq2}, for different values of $\nu_r^\prime$ and its steady state value estimated. We found that $\nu_c^\prime$ is independent of the exact value of $\nu_r^\prime$ as long as it is non-zero, but different from the $\nu_r^{\prime}=0$ value. The results are shown in Fig.\ref{fig:fig2}, for cases studied in sections A and B.

\vspace{0.7cm}

\subsection{Comparison with the results in Antal et al. \cite{antal} and Padinhateeri et al. \cite{ranjith}}

The models studied in Antal et al. \cite{antal} and Padinhateeri et al. \cite{ranjith} have several similarities with the model studied in this paper, therefore we make a brief comparative study of the results in this section. Since both are effective one-dimensional models for the microtubule, it is apt to compare our results with theirs at the protofilament level. A comparison between the notations for various quantities in our model and those in Ref. \cite{antal} and Ref. \cite{ranjith} is given in Table \ref{tab:tab2}.

In  Ref. \cite{antal}, the authors study a one-dimensional model, with growth, hydrolysis (spontaneous), D-mer dissociation from the tip and protofilament rescue by addition of a T-mer to a D-tipped filament. In addition to the probability of catastrophe, these authors also studied quantities such as length distributions for the cap and GTP/GDP islands in the filament. Here, we restrict the discussion to their result for the probability of a ``global'' catastrophe between two growth events. In the limit of infinitely fast D-mer detachment (which rules out rescues), and for $k_g\gg r$, this probability is shown to have the form (in our notation)

\begin{equation}
P_{\mathrm{cat}}\equiv \frac{\nu_c^{\prime}}{k_g}=\frac{r}{r+k_g}\prod_{n=1}^{L-1}(1-e^{-nr/k_g}),
\label{eq:eqR3}
\end{equation}
where $L$ is the total number of monomer units in the filament. The product in the above expression gives the probability that except for the unit at the tip, all other monomer units are D-mers and therefore, a catastrophe event will result in the entire filament vanishing by depolymerization (hence the name ``global" catastrophe). In our model, as in other similar models \cite{FHL,ranjith}, we use a much less restrictive definition of catastrophe, according to which the occurrence of at least two D-mers at the tip would initiate catastrophe. Then, by considering only the first term (with $n=1$) in the above product, and after using the approximation $1-e^{-r/k_g}\simeq r/k_g$, the protofilament catastrophe predicted by Eq.\ref{eq:eqR3} is seen to be exactly the same as Eq.\ref{eq:eq27} in this paper.

\begin{table}
\begin{tabular}{|c|c|c|c|}
\hline

Rate for & Our model &  Ref. \cite{antal}&Ref. \cite{ranjith} \\
\hline
Monomer addition (T$\to$ T) & $k_g$ &  $\lambda$ & $U$\\
\hline
Monomer addition (T$\to$ D) & $\nu_r^{\prime}$ &$ p\lambda$& $U$ \\
\hline
Monomer dissociation (T) & $k_d$   &absent& $W_T$ \\
\hline
Monomer dissociation (D) & absent & $\mu$ &$W_D$\\
\hline
Spontaneous hydrolysis & $r$    &1& $r$ \\
\hline
Induced hydrolysis & $k_h^*$ & absent&absent \\
\hline
Catastrophe frequency & $\nu_c^{\prime}$ &$C(\lambda)k_g$& $f_c(1)-r$ \\
\hline
\end{tabular}
\caption{The table gives a comparison of the notations used in our paper, and those in Ref. \cite{antal} and Ref. \cite{ranjith} . The additional factor $-r$ in the last entry of last column is required to match the definitions of catastrophe in both models (see discussion in text). All the rates except that corresponding to \cite{antal} are expressed in units of s$^{-1}$ whereas as  those in \cite{antal} are dimensionless.}
\label{tab:tab2}
\end{table}

The model in Ref. \cite{ranjith}, while in many ways identical with our model, also introduces a parameter $N$ which is the minimum number of D-mers required to be present at the tip in order 
for catastrophe to occur. The second term in the expression for catastrophe (see Eq.6 in Ref. \cite{ranjith}) is absent in our model, since, in this work (following the definition in Ref. \cite{FHL}), catastrophe is defined as the transition from single GTP-tipped state (probability $P_1$) to a GDP-tipped state (probability $P_0$).  In their model, the expression for catastrophe frequency takes the following form, in our notation (Eq.5 in the supplementary material of Ref. \cite{ranjith}, with the substitution $N=1$ to make it consistent with our model):

\begin{equation}
\nu_c^{\prime}=k_h(1-bq)~~;~~~b=\frac{k_g-q(k_h+r)}{k_g-k_h q},
\label{eq:eqR1}
\end{equation}
where, $q$, the probability that the (one-dimensional) filament is GTP-tipped, is shown to obey a cubic equation (Eq.29 in Ref. \cite{ranjith2010}) within a mean-field approximation. (For comparison, the catastrophe frequency defined in Ref. \cite{ranjith} is $f_c(1)=\nu_c^{\prime}+r$, the last term added to take into account catastrophe events where the only the terminal GTP unit undergoes hydrolysis. In this paper, we adopt the view that such events do not constitute observable catastrophes, as such a filament is almost immediately rescued by the exposure of the underlying GTP-region upon loss of the terminal GDP unit). In the limit of large $k_g$ and with $W_D=0$ (the rate of dissociation of D-mers from the filament, which is absent in our model), the solution to the cubic equation can be shown to be 

\begin{equation}
q\simeq 1-\frac{r}{k_g}~~~~k_g\gg k_h,r.
\label{eq:eqR2}
\end{equation}

Substitution of Eq.\ref{eq:eqR2} into Eq.\ref{eq:eqR1} leads to Eq.\ref{eq:eqXX4} with $\Gamma=2rk_h$, i.e., the asymptotic expressions match between the models, 
as expected. 

\section{Results II: Microtubule catastrophe}

\subsection{Relation between microtubule and protofilament catastrophes}

Although microtubule catastrophes are regularly observed both {\it in vitro} and {\it in vivo}, there is still some lack of clarity about the sequence of events in a growing microtubule that ultimately leads to its transition to shrinking state (see \cite{anderson} for a recent review). It has been shown experimentally, first by Odde et al. \cite{odde} and more recently by Stepanova et al. \cite{stepanova} and Gardner et al. \cite{gardner} that microtubule catastrophes have a certain history-dependence, i.e., the probability for a microtubule to undergo catastrophe appears to depend on how long its has been growing. Based on these observations, it was suggested that catastrophe is a multi-step process, which requires a certain number of events to occur for its materialization. The precise nature of these events is still speculative, but one suggested possibility \cite{gardner}  is that a minimum number of protofilaments are required to lose their GTP-tips before the entire fi
 lament can undergo catastrophe. Here, we subject this conjecture to test by modeling catastrophe as a first-passage event that will occur once the system of protofilaments reaches a state where a minimum number of them have lost their GTP-tips. 

Our working postulate shall be that global (microtubule) catastrophe ensues when the set of 13 protofilaments reaches a state where a certain number $n^*$ exist in GDP-tipped state (a similar conjecture was also used in Ref. \cite{nedelec}). We then performed numerical simulations of catastrophes using a Gillespie algorithm \cite{Gill} for different values of $n^*$, based on the above conjecture. 

\subsection{Numerical simulations}

In the simulations, every microtubule is a set of 13 protofilaments, all evolving independently with time. The dynamics of each protofilament involves growth, hydrolysis (both spontaneous and vectorial, although the latter could also be interpreted as T-dimer detachment from the tip), catastrophe and rescue (both at the protofilament level, as discussed in the last section). Each protofilament evolves in variable size time jumps until a maximum time of $T_{\max}$ (which varies from 400s for small $k_g$ up to 7000s for large $k_g$) is crossed. It is assumed that, even after undergoing catastrophe, a protofilament will not start shrinking immediately, but will stop growing (i.e., is ``paused") until a rescue occurs.  

In the next stage of the simulation, the states (whether growing or paused due to a catastrophe) of 13 protofilaments belonging to one microtubule are inspected as a function of time, starting from $t=0$ until $t=T_{\max}$, in steps of $\delta t$. In this way, a catastrophe time for a microtubule is determined, to a precision of $\delta t$, as the first instant when at least $n^*$ of its constituent protofilaments are found to exist in a paused state due to catastrophe. Although it is possible in principle to choose $\delta t$ to be as small as possible, we make a specific choice of $\delta t=k_h^{-1}$ since it is reasonable to assume $k_h\sim k_d$, the T-mer detachment rate (see Sec.IIIA), and therefore the time scale of initiation of catastrophe may be roughly taken to be its inverse (assuming that D-mers also detach at a similar rate, post-catastrophe). In the simulations, we will assume that a microtubule, having undergone catastrophe is never rescued; hence once the cata
 strophe time has been identified for a certain microtubule, we count the filament as `dead' and move on to the next microtubule. In the last stage, the number of microtubules $N(t)$ which are  still `alive' at time $t$ is computed as a function of $t$, and the survival probability is defined as $P_s(t)=N(t)/N(0)$. The effective microtubule catastrophe is then evaluated as $\nu_c(t)=-\dot{P_s}/P_s$, where $\dot{P_s}$ denotes the time derivative. The steady state regime for $\nu_c(t)$ is then identified visually, and its average and standard deviation computed. 

\begin{figure}[h]
 \centering
 \includegraphics[scale=.35,keepaspectratio=true]{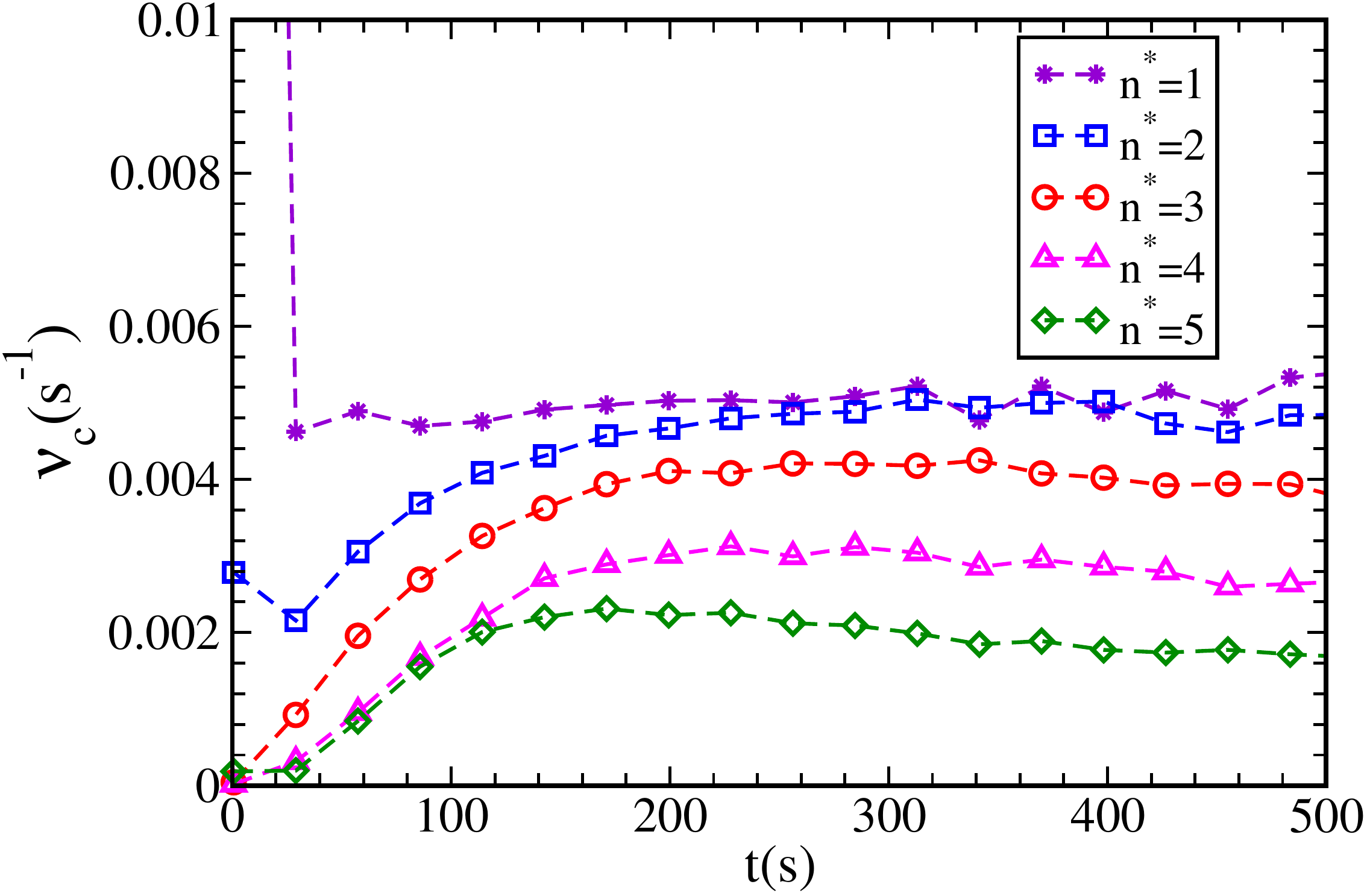}
 \caption{(color online)The figure shows the time evolution of microtubule catastrophe for four different values of $n^*$. The respective values of $r$ for $n^*=1,2,3,4,5$ are 0.004s$^{-1}$, 0.008s$^{-1}$, 0.013s$^{-1}$, 0.018s$^{-1}$, 0.03s$^{-1}$ (see explanation in text). The other parameters are $k_h$=7s$^{-1}$ and $\nu_r^{\prime}=10^{-3}$s$^{-1}$.}
 \label{fig:fig4}
\end{figure}

In the first set of simulations, we looked at the dynamics of microtubule catastrophe where we obtained  time evolution curve for catastrophe  for different $n^*$ and compared our results with experimental studies by Gardner et al. \cite{gardner}. Since microtubule rescue was not observed in these experiments, we chose a sufficiently small value, i.e., $\nu_r^{\prime}=10^{-3}$s$^{-1}$ for the protofilament rescue in the simulations. In principle, $\nu_r^{\prime}$ could depend on $k_g$, but experiments \cite{walker} show that the dependence of rescue on tubulin concentration and growth rate is much weaker than that of catastrophe, and is effectively absent at large $k_g$. The rate of vectorial hydrolysis $k_h^*$ has never been measured experimentally in microtubules; however, as we showed in Sec.III, this parameter enters our equations in the combination $k_h=k_h^*+k_d$, where $k_d$ is the rate of dissociation of a protofilament-incorporated T-mer. Therefore, we may use the av
 ailable estimates of $k_d$ as a measure of $k_h$. Margolin et al. \cite{alber} estimated that $k_d\sim 1-10$s$^{-1}$, by studying the experimental data of Gardner et al. \cite{gardner}, depending on the tubulin concentration. It turned out that a high $k_h$ value is necessary to reproduce the experimental data of \cite{gardner}, and therefore, we fixed $k_h$=7s$^{-1}$ and tuned $r$, for each $n^*$, such that the steady state value of microtubule catastrophe is comparable to the experimental value 0.005s$^{-1}$ for $k_g=24.375\mathrm{s}^{-1}$ (corresponding to a growth velocity $v_g=15\mathrm{nms}^{-1}$ in the experiment, with the simple conversion formula $k_g=v_g/\delta$ with $\delta=8\mathrm{nm}/13=0.6\mathrm{nm}$).
 The time evolution curves thus obtained are shown in Fig.\ref{fig:fig4} for $n^*=1,2,3,4$ and 5. As discussed in the next section, mathematical calculations using an exactly  soluble ``equilibrium" model show that $\nu_c$ is bound from above for fixed $\nu_r^{\prime}$ and $k_h$, and is insensitive to increase in $r$ beyond a point. This explains the logic behind different steady state values corresponding to different $n^*$. From Fig.\ref{fig:fig4}, it is evident that the time evolution curve of $n^*=1$ is too flat compared to the experimental curve, whereas  $n^*=2$ and 3 appear similar, reaching saturation in about 300s (compared to $\sim 400$s in experiments). 

\begin{figure}[h]
 \centering
 \includegraphics[scale=.31,keepaspectratio=true]{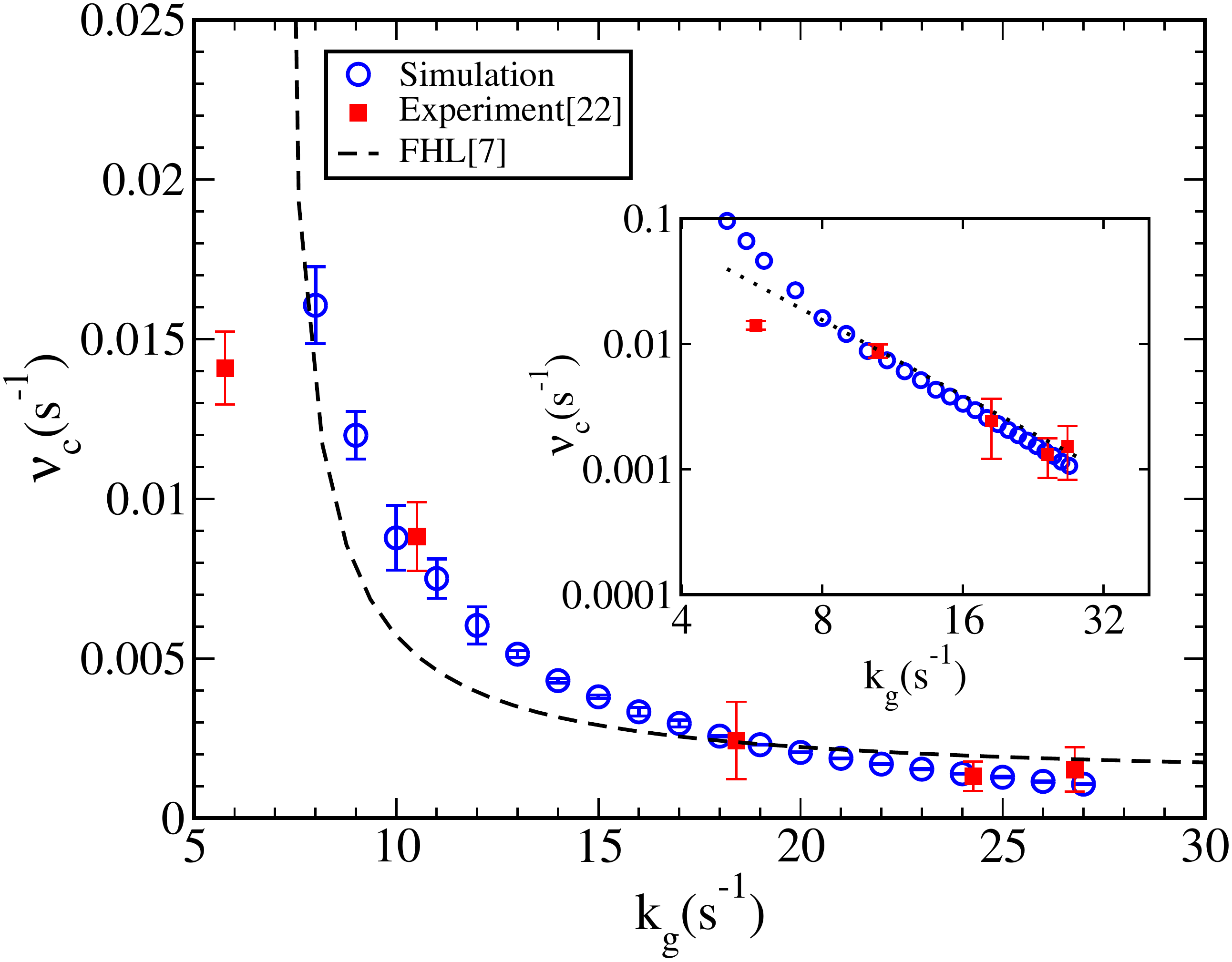}
 \caption{(color online)The figure shows a comparison between the experimental data for catastrophe versus growth rate reported in Ref. \cite{drechsel}, and numerical simulation using a Gillespie algorithm, with $n^*=2$. For the simulation data shown in the figure, the value of the parameters are $k_h=4\mathrm{s}^{-1}$, $r=7\times10^{-3}\mathrm{s}^{-1}$ and $\nu_r^{\prime}$=$10^{-3}$s$^{-1}$. The dashed line shows the prediction from the FHL model \cite{FHL}, reproduced in Eq.\ref{eq:eq14}. Inset: Simulation and experimental data plotted against $k_g$ on a double logarithmic scale. A dotted line with slope $-2$ is also shown for comparison  
(see also  discussions following Eq.\ref{eq:eqM8} in the next subsection).}
 \label{fig:fig6}
\end{figure}

For further analysis, we chose both $n^*=2$ and $n^*=3$. Next, we studied the behavior of steady state catastrophe frequency, as function  of growth rate. The steady state catastrophe is calculated from the survival probability as discussed above. We take the experimental data presented in Ref. \cite{drechsel} for reference. Initially, both $\nu_r^{\prime}$ and $k_h$ are fixed at the previous values used in the kinetic analysis. For $k_g=26.78$s$^{-1}$ (the largest value in Ref.  \cite{drechsel}), we then tuned $r$ so that the simulation result for $\nu_c$ in steady state agrees with the experiment. In the next step, we simulated lower $k_g$ values for the same set of parameters and compared with the experimental data. To improve the agreement, both $k_h$ and $r$ were tuned in successive steps. The best fit with the experimental data, determined by visual inspection, was obtained for $k_h=4$s$^{-1}$ and $r=7\times 10^{-3}$s$^{-1}$, and is shown in Fig.\ref{fig:fig6} and Fig.\
 ref{fig:fig6+}, for $n^*=2$ and $n^*=3$ respectively. It is clear from the figures, and the logarithmic scale versions in the insets, that $n^*=2$ (Fig.\ref{fig:fig6}) gives a better overall fit compared to 3 (except at very small $k_g$, where the reverse seems to be true). The FHL result (Eq.\ref{eq:eq14}) is also shown for comparison in the figures. Note that both simulation as well as experimental data appear to follow a power-law decay $\nu_c\propto k_g^{-2}$ asymptotically,  in contrast to the FHL result, which predicts a non-zero asymptotic value as $k_g\to\infty$.  More discussion on the large $k_g$ asymptotics is given in the next section, where we discuss analytical results.

 A few sample curves showing survival probability as a function of time, for $n^*=3$ at different $k_g$ values are shown in Fig.\ref{fig:fig5}. We also attempted simulations with $k_h=0$ and tuning only $r$. Following the above procedure, for the largest value of $k_g$, the estimated value of $r$ was greater than 10s$^{-1}$, much higher than other known estimates (see Table IV), but even with this value, we were not successful in reproducing the experimental data for the lower $k_g$ values. 

\vspace{0.5cm}

\begin{figure}[h]
 \centering
 \includegraphics[scale=.32,keepaspectratio=true]{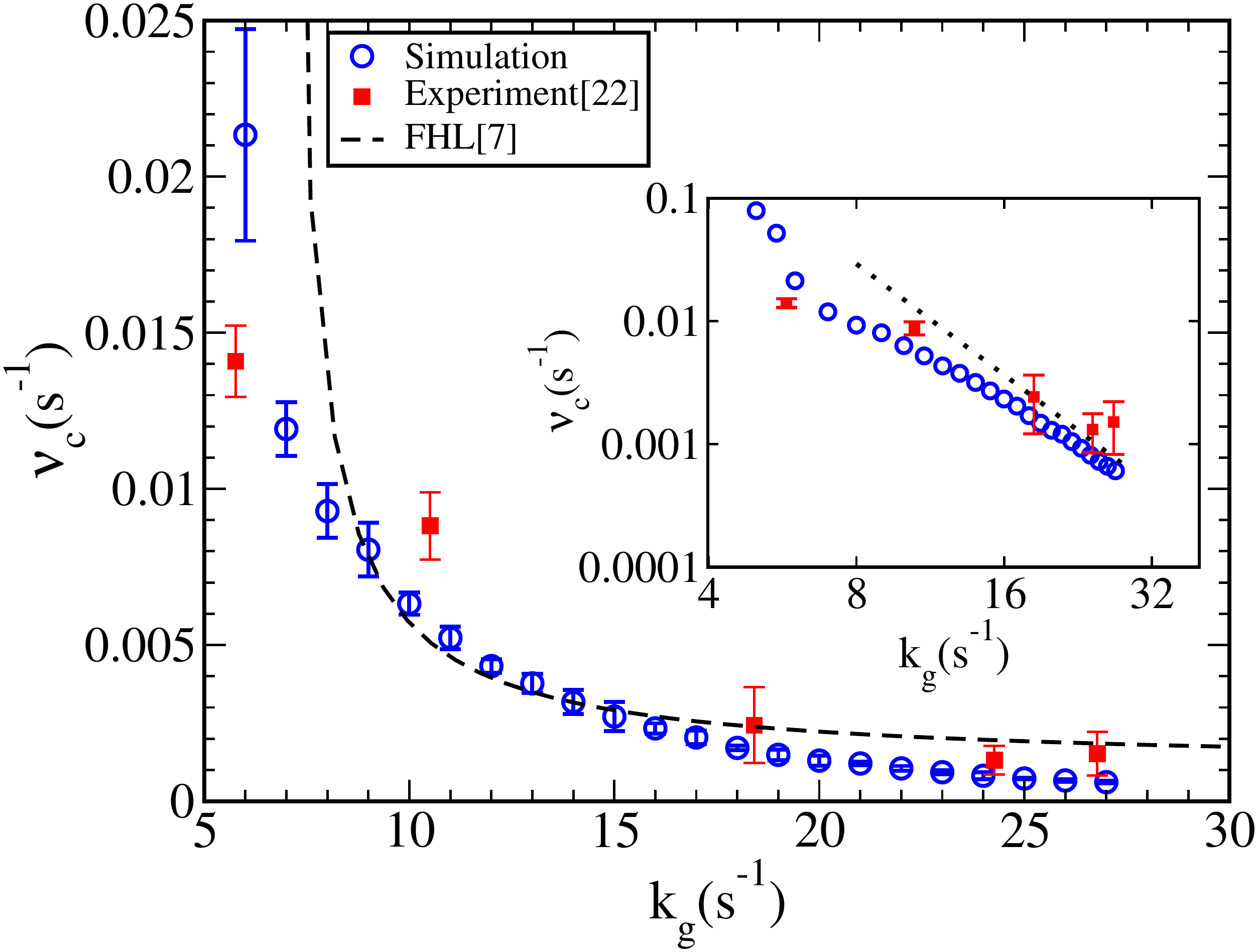}
 \caption{(color online) Similar to Fig.\ref{fig:fig6}, but with $n^*=3$. All other parameters have the same values as in the previous figure. Inset: the simulation data shown on a double logarithmic scale against $k_g$. The dotted line has slope $-3$ (see also  discussions following Eq.\ref{eq:eqM8} in the next subsection).}
 \label{fig:fig6+}
\end{figure}

\begin{figure}[h]
 \centering
 \includegraphics[scale=.31,keepaspectratio=true]{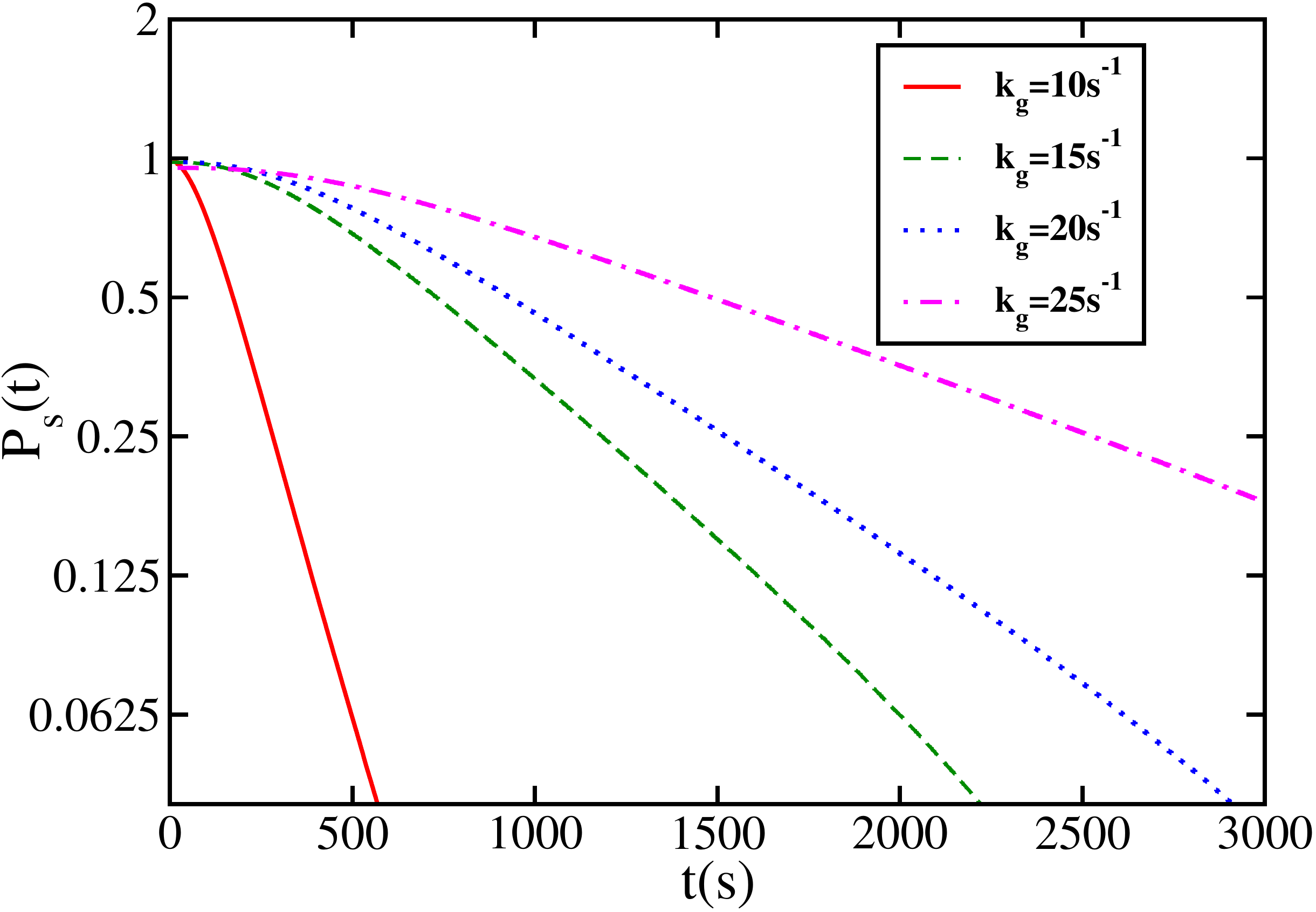}
 \caption{(color online)The figure shows survival probability of microtubules, plotted as a function of time in a semi-logarithmic scale at various growth rates. The curves shown here correspond to the steady state data displayed in Fig.\ref{fig:fig6+}.}
 \label{fig:fig5}
\end{figure}

\subsection{Analytical results}

As we saw in the last subsection, stochastic numerical simulations give results which agree well with experimental data, both kinetic and steady state, and therefore provides support for the conjectured relation between protofilament and microtubule catastrophes outlined in the beginning of this section. However, a precise quantitative relation between the two quantities is still lacking, in the absence of which the mathematical results of the last section is of limited use in analyzing biological data of microtubule catastrophes. In this section, we introduce and study two models for this purpose. In the first model, we derive an analytical expression for the microtubule catastrophe under steady state conditions, where microtubule rescue is allowed. Under these assumptions, this model will be shown to satisfy a detailed balance condition, and for this reason (and this alone) we shall refer to it as the ``equilibrium" model in a limited sense.  The second model will be called
  a  ``quasi-equilibrium" model, where all protofilament-related quantities are assumed to have reached their steady state values as earlier, but microtubule catastrophe is assumed to be a slow and irreversible process. Comparison with the simulation results from the last subsection shows that predictions of both the models consistently overestimate the catastrophe frequency (compared to simulations), but the second model gives comparatively better results. \\

(i) {\it ``Equilibrium" model} (reversible catastrophe):\\

In this model, we will assume that each protofilament is found in one of the two states, growing  or non-growing, depending on whether it is GTP-tipped or not. The 13 protofilaments belonging to a microtubule evolve independently with time. A protofilament undergoes catastrophe only if the GTP cap is lost through a $1\to 0$ transition, whereas all the $n\to 0$ transitions with $n\geq 2$ are not counted as catastrophes. When a microtubule first reaches a configuration where $n^*$ protofilaments have undergone catastrophe and remain in that state (we shall refer to such a protofilament as existing in a ``catastrophed" state), it is defined to instantaneously undergo the catastrophe transition; however, no tubulin dissociation or disintegration of the microtubule lattice is assumed to take place. Instead, each protofilament goes through its own dynamics independently of the others. In this sense, the catastrophe transition in this model is a first passage event and is reversible
 ; as soon as the system enters a configuration with $14-n^*$ protofilaments in ``non-catastrophed" state, the filament itself is assumed to be ``rescued''. Although certainly not a very accurate description of the real microtubule dynamics, this model is exactly soluble and provides a mathematical expression for the catastrophe frequency in terms of protofilament catastrophe frequency and other parameters.

The probability that a certain protofilament is in a ``catastrophed" state is $P_0^{\prime}<P_0$. From Eq.\ref{eq:eq1}, it is seen that $P_0^{\prime}$ satisfies the equation

\begin{equation}
 \frac{\partial P_0^{\prime}(t)}{\partial t} =(r+k_h)P_1(t)-\nu_r^{\prime} P_0^{\prime}(t).
\label{eq:eqM1}
\end{equation}

Let us now denote by $Q_n(t)$ the probability that the microtubule has $n$ protofilaments which have undergone catastrophe at some time $t^{\prime}<t$ and not rescued until time $t$. It then follows that 

\begin{equation}
Q_n(t)={13 \choose n}\left(P_0^{\prime}\right)^{n}\left(1-P_0^{\prime}\right)^{13-n}.
\label{eq:eqM2}
\end{equation}

Using Eq.\ref{eq:eqM1} and Eq.\ref{eq:eqM2}, the time evolution of $Q_n(t)$ may be conveniently expressed in the form 

\begin{equation}
\frac{dQ_n}{dt}=k_{-}^{n+1}Q_{n+1}+k_{+}^{n-1}Q_{n-1}-(k_{+}^{n}+k_{-}^{n})Q_n,
\label{eq:eqM3}
\end{equation}
where the `rates' are defined by 

\begin{equation}
k_+^{n}=(13-n)\nu_c^{\prime\prime}~~;~~k_{-}^{n}=n\nu_r^{\prime} 
\label{eq:eqM4}
\end{equation}
and 

\begin{equation}
\nu_c^{\prime\prime}(t)\equiv \nu_c^{\prime}(t)\frac{1-P_0(t)}{1-P_0^{\prime}(t)}
\label{eq:eqM5}
\end{equation}
is an `effective' protofilament catastrophe frequency. It is easily seen from Eq.\ref{eq:eqM3} that in the long-time limit, the quantities $Q_n(t)\to Q_n$, their `steady state' values, which satisfy detailed balance, i.e., 

\begin{equation}
\frac{Q_{n+1}}{Q_n}=\frac{k_n^{+}}{k_{n+1}^{-}}.
\label{eq:eqM5+}
\end{equation}

For a certain $n^*$, then, the microtubule catastrophe is given by

\begin{equation}
\nu_c(t)=\frac{(14-n^*)\nu_c^{\prime\prime}(t)Q_{n^*-1}(t)}{\sum_{0}^{n^*-1}Q_j(t)}.
\label{eq:eqM6}
\end{equation}

After substitution of the steady state probabilities

\begin{equation}
P_{0}=\frac{\nu_c^{\prime}k_h+r(r+k_h)}{\nu_c^{\prime}k_h+(r+\nu_r^{\prime})(r+k_h)}~~;~~P_0^{\prime}=\frac{\nu_c^{\prime}}{\nu_r^{\prime}}(1-P_0),
\label{eq:eqM7}
\end{equation}
obtained from Eq.\ref{eq:eq1}, Eq.\ref{eq:eq2} and Eq.\ref{eq:eqM1} into Eq.\ref{eq:eqM2} and Eq.\ref{eq:eqM5}, we arrive at the result:

\begin{equation}
\nu_c^{\prime\prime}=\frac{\nu_c^{\prime}\nu_r^{\prime}(r+k_h)}{(r+\nu_r^{\prime})(r+k_h)-r\nu_c^{\prime}}.
\label{eq:eqM8}
\end{equation}

For large values of $k_g$, $\nu_c^{\prime}\ll r+k_h$. In this limit, we get the simplified and useful expression 
 
\begin{equation}
\nu_c^{\prime\prime}\simeq \frac{\nu_c'\nu_r'}{(r+\nu_r')}~~~~~k_g\to\infty.
\label{eq:eqM9}
\end{equation}

The following points are of interest here. While $\nu_c^{\prime\prime}\propto \nu_c^{\prime}\propto k_g^{-1}$ as $k_g\to\infty$, the prefactor also becomes independent of $r$ if $r\gg \nu_r^{\prime}$, when $k_h>0$ (see Eq.\ref{eq:eqXX4} and Table \ref{tab:tab1}). In this limit, it is $k_h$ that primarily determines $\nu_c^{\prime\prime}$ (unless, of course, $k_h=0$, in which case $r$ assumes this role). 

The microtubule catastrophe, under these assumptions, for $n^*=2$ and 3,  are given by 

\begin{eqnarray}
\nu_c^{\mathrm{eq}}=\nu_c^{\prime\prime}\frac{156\eta}{1+13\eta}~~~~~n^*=2\nonumber\\
\nu_c^{\mathrm{eq}}=\nu_c^{\prime\prime}\frac{858\eta^2}{1+13\eta+78\eta^2}~~~~n^*=3,
\label{eq:eqM8}
\end{eqnarray}
where $\eta=\nu_c^{\prime \prime}/\nu_r^{\prime}$. Using Eq.\ref{eq:eqM5+} and Eq.\ref{eq:eqM6}, it is easily seen that $\nu_c^{\mathrm{eq}}\simeq (14-n^*)\nu_c^{\prime\prime}$ when $\eta\gg 1$, while for $\eta\ll 1$ (equivalent to large values of $k_g$), the following limiting behaviors are observed:

\begin{eqnarray}
\nu_c^{\mathrm{eq}}\simeq 156\frac{\nu_c^{{\prime\prime}^2}}{\nu_r^{\prime}}~~~n^*=2, k_g\to\infty\nonumber\\
\nu_c^{\mathrm{eq}}\simeq 858\frac{\nu_c^{{\prime\prime}^3}}{\nu_r^{\prime 2}}~~~n^*=3, k_g\to\infty.
\label{eq:eqM9}
\end{eqnarray}

In particular, Eq.\ref{eq:eqM9} predicts that $\nu_c\sim k_g^{-2}$ and $k_g^{-3}$ asymptotically, for large $k_g$, for $n^*=2$ and $n^*=3$ respectively. This asymptotic behavior agrees with our simulation results shown in Fig.\ref{fig:fig6} and Fig.\ref{fig:fig6+}.  The experimental data of Drechsel et al. \cite{drechsel} too appear to follow a power law decay with en effective exponent close to 2 (Fig.\ref{fig:fig6}), which would suggest that catastrophe in two protofilaments would initiate the filament catastrophe.  However, more recent experimental data reported by Janson et al. \cite{janson} and analyzed in detail by Brun et al. \cite{nedelec} suggest that $\nu_c\sim k_g^{-1}$ at large $k_g$, which apparently seems to be consistent with choosing $n^*=1$ in our model; however, the kinetic data in this case (Fig.\ref{fig:fig4}) do not agree with experiments in Gardner et al. \cite{gardner}.\\

(ii) {\it ``Quasi-equilibrium" model} (irreversible catastrophe):\\

The experiments reported in neither \cite{drechsel} nor \cite{gardner} observed microtubule rescue. The Gillespie simulations discussed in Sec. V B were designed to be consistent with this observation, but the equilibrium model is not. This discrepancy may be partly rectified without significant additional effort, by using the steady state rates in Eq.\ref{eq:eqM3} to construct a ``quasi-equilibrium" model, where microtubule rescue is prohibited, and, the state $n=n^*$ becomes absorbing (naturally, the detailed balance condition in Eq.\ref{eq:eqM5+} no longer applies). In this approximation (where the transition rates $k_{\pm}^n$ are assumed to have the same values as the equilibrium model; hence the quasi-equilibrium nature of the model), the dynamical equations in Eq.\ref{eq:eqM3} takes the form $\dot {\bf Q}=\Lambda_{n^*} {\bf Q}$, where ${\bf Q}$ is the column vector ${\bf Q}=[Q_0~ Q_1~Q_2~...Q_{n^*-1}]^{\mathrm{T}}$ and $\Lambda_{n^*}$ is an $n^*\times n^*$ matrix. For $
 n^*=2$ and $n^*=3$, $\Lambda_{n^*}$ is given by

\begin{eqnarray}
\Lambda_2=\left[ \begin{array}{cc}
-13\nu_c^{\prime\prime} & \nu_r^{\prime}\\
13\nu_c^{\prime\prime} & -(\nu_r^{\prime}+12\nu_c^{\prime\prime})
\end{array}\right]
\nonumber\\
\Lambda_3=\left[ \begin{array}{ccc}
-13\nu_c^{\prime\prime} & \nu_r^{\prime} & 0\\
13\nu_c^{\prime\prime} & -(\nu_r^{\prime}+12\nu_c^{\prime\prime}) & 2\nu_r^{\prime}\\
0 & 12\nu_c^{\prime\prime} & -(2\nu_r^{\prime}+11\nu_c^{\prime\prime})
\end{array}\right],
\label{eq:eqM10}
\end{eqnarray}
whose eigenvalues determine the time evolution of ${\bf Q}$. The definition of microtubule catastrophe frequency remains the same as in Eq.\ref{eq:eqM6}, but now, both the numerator and the denominator become time dependent, and decay with time. However, in the long-time limit, their ratio becomes a constant, and is given by the largest (negative) eigenvalue of $\Lambda$, which we denote by $-\lambda_{\min}$. To prove this assertion, we put $Q_n(t)\sim a_n\exp(-\lambda_{\min} t)$ as $t\to\infty$. By consecutive substitutions in the above vector equation, with $\Lambda$ given by Eq.\ref{eq:eqM10}, the relation $(11\nu_c^{\prime\prime}-\lambda_{\min})a_2=\lambda_{\min}(a_0+a_1)$ can be shown to hold true for $n^*=3$, while another, similar relation $12\nu_c^{\prime\prime}a_1=\lambda_{\min}(a_0+a_1)$ is valid for $n^*=2$. Substitution of these relations in Eq.\ref{eq:eqM6} shows that $\nu_c=\lambda_{\min}$ for both these cases, in the present model. For $n^*=2$, $\lambda_{\min}$
  is the solution of a quadratic equation, and is given by

\begin{equation}
\lambda_{\min}=\frac{\nu_r^{\prime}+25\nu_c^{\prime\prime}-\sqrt{\nu_r^{\prime 2}+50\nu_r^{\prime}\nu_c^{\prime\prime}+\nu_c^{\prime\prime 2}}}{2}.
\label{eq:lambda}
\end{equation}

In the large $k_g$ limit, we may assume $\nu_c^{\prime\prime}\ll \nu_r^{\prime}$. In this case, a binomial expansion of the square root in the above equation yields $\lambda_{\min}\simeq 156\nu_c^{\prime\prime 2}/\nu_r^{\prime}$, which matches the corresponding equilibrium model result in Eq.\ref{eq:eqM9}. For $n^*=3$, the eigenvalue satisfies a cubic equation and hence $\lambda_{\min}$ was found by diagonalization of the matrix $\Lambda$ using {\it Mathematica} \cite{math}. 

\begin{figure}[h]
 \centering
 \includegraphics[scale=.29,keepaspectratio=true]{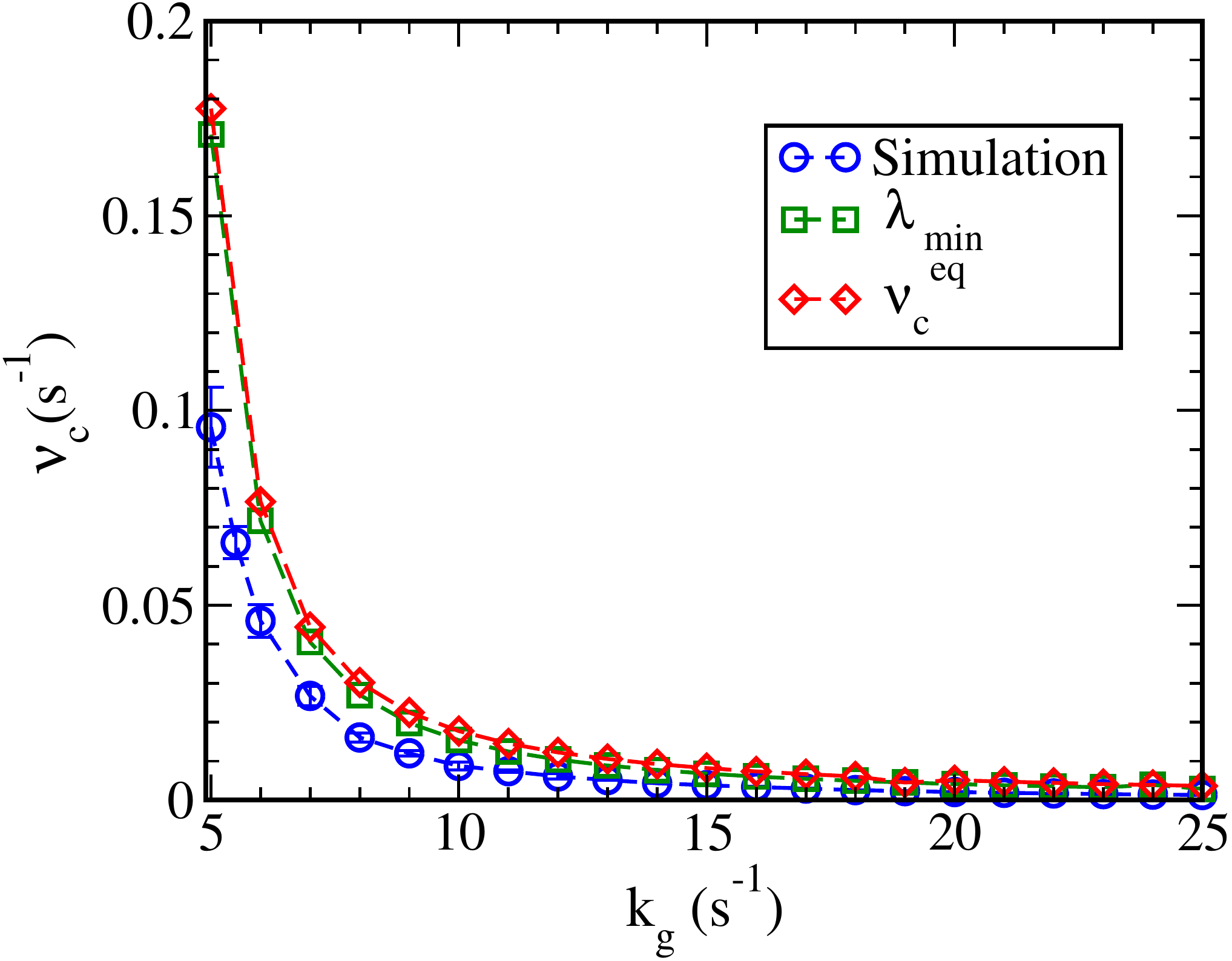}
 \caption{(color online)The figure shows a comparison between $\nu_c$ measured in simulations and the analytical results in Eq.\ref{eq:eqM8} and Eq.\ref{eq:lambda}, for $n^*=2$.}
 \label{fig:fig7}
\end{figure}

Comparisons of the analytical results in Eq.\ref{eq:eqM8} and Eq.\ref{eq:lambda} with the corresponding numerical data are shown in Fig.\ref{fig:fig7} ($n^*=2$) and Table \ref{tab:tab3} ($n^*=3$). 

\begin{table}
 \begin{tabular}{|c|c|c|c|}
  \hline
$k_g (\mathrm{s}^{-1})$ ~& ~Simulation (s$^{-1}$) & ~$\lambda_{\min}$ (s$^{-1}$) & ~ $\nu_c^{\mathrm{eq}}$ (s$^{-1}$) \\
\hline
5& 7.9$\pm0.4\times10^{-2}$ & 1.5$\times10^{-1}$& 1.6$\times10^{-1}$\\
\hline
10&6.3 $\pm0.35\times10^{-3}$ & 1.2$\times10^{-2}$    &1.5$\times10^{-2}$     \\
\hline
15& 2.7 $\pm0.47\times10^{-3}$ & 4.8 $\times10^{-3}$   &6.6$\times10^{-3}$     \\
\hline
20&  1.3 $\pm 0.17\times10^{-3}$  &  2.7  $\times10^{-3}$ & 3.9 $\times10^{-3}$   \\
\hline
25&   7.2 $\pm 0.18\times10^{-4}$ & 1.7$\times10^{-3}$  & 2.6$\times10^{-3}$       \\
\hline
 \end{tabular}
\caption{The table gives a comparison of the steady state catastrophe frequency as measured in simulations for a few chosen values of $k_g$, versus the predictions of the quasi-equilibrium and equilibrium models respectively, for $n^*=3$.}
\label{tab:tab3}
\end{table}

\begin{table}
\begin{tabular}{|c|c|c|}
\hline
Estimated $r$ & Reference & Year\\
\hline
2.2$\times 10^{-3}$   & Flyvbjerg et al. \cite{FHL} & 1996\\
\hline
0.95 & VanBuren et al. \cite{vanburen} & 2002\\
\hline
0.029 & Brun et al. \cite{nedelec} & 2009\\
\hline
0.3   &   Piette et al. \cite{piette} & 2009\\
\hline
0.25  &   Padinhateeri et al. \cite{ranjith} & 2012\\
\hline
0.7 &    Margolin et al. \cite{alber} & 2012\\
\hline
7$\times10^{-3}$ & this work ($k_h=4\mathrm{s}^{-1}$) & 2013\\
\hline
\end{tabular}
\caption{ The table gives a comparison of the estimates of the spontaneous hydrolysis rate obtained by different authors. 
In  Ref. \cite{FHL}, the rate $r^{\prime}$ per monomer unit length $\delta x$ was given, which was converted as $r=r^{\prime}\delta x$, with $\delta x=0.6$nm. 
All the $r$ values are in units of s$^{-1}$. }
\label{tab:tab4}
\end{table}

\section{Conclusions}

In the present paper, we have studied the stochastic model of GTP cap dynamics, introduced by Flyvbjerg, Holy and Leibler (FHL) \cite{FHL}. Both spontaneous and vectorial hydrolysis have been included in the model, partly because the effect of the latter in the associated dynamical equation is the same as that of a dissociation term for a filament-incorporated GTP-dimer. Unlike the FHL study, we employ a discrete formalism here, which is more appropriate for individual protofilaments. We do not make use of an effective one-dimensional picture of a microtubule unlike many previous authors \cite{FHL,antal,ranjith}; rather, we define events of catastrophe and rescue for each protofilament, which are then related to microtubule catastrophe via first passage concepts. The protofilament catastrophe and rescue events are defined analogous to the corresponding events in an entire microtubule filament; catastrophe here refers to the loss of a GTP-dimer tip in the protofilament, wherea
 s rescue refers to the addition of a GTP-dimer to a GDP-tipped protofilament.

We find that even at the level of a single protofilament (i.e., a one-dimensional filament), the predictions of the model, in general, differ from the predictions of the corresponding continuum model of FHL. We also considered both steady state (protofilament rescue present) and non-steady state (protofilament rescue absent) situations; the distinction between these has never been clearly addressed in the literature (for example, in the FHL model, the loss of GTP cap is an absorbing state which cannot be rescued whereas in the models used in \cite{antal,ranjith}, a GTP-dimer may attach to a GDP-tipped filament with non-zero probability and `rescue' it). We show rigorously that while catastrophe frequency in the stochastic model is independent of the protofilament rescue rate, it depends, nevertheless, on whether steady state or non-steady state conditions are employed. 

In spite of these differences, we show that the asymptotic behavior of protofilament catastrophe in the limit of large values of protofilament growth rate $k_g$ is simply $\nu_c^{\prime}\sim \Gamma k_g^{-1}$, where the proportionality constant depends on the specific conditions used. This remarkable universal property perhaps partly explains why predictions of many different models have been found to fit well with available experimental data (e.g., \cite{drechsel}). We also compared predictions of our model with the exact results in Antal et al. \cite{antal}, as well as mean-field model of Padinhateeri et al. \cite{ranjith}, and found that the mathematical results of the models agree in the asymptotic regime discussed above, under appropriate conditions. A comparison of our estimate of $r$, with those arrived at by other authors is illuminating. We observe that while there is a reasonable agreement between the values predicted by purely kinetic studies (except Ref. \cite{ranj
 ith}), they differ significantly from models where the energetics of the microtubule filament is explicitly included, which typically predict much higher values of $r$. However, since both types of models are seemingly able to demonstrate agreement with experiments, it would be interesting to know if the rate used in the kinetic models should be regarded as an effective parameter which hides some information about the energetics of binding between tubulin dimers, both within a protofilament and between protofilaments. We hope that our study will stimulate further investigations in this direction.

Our model also predicts that 2-3 protofilaments out of thirteen are required to lose the GTP cap to initialize a catastrophe event, based on an analysis of the (microtubule) age dependence of catastrophes, observed in experiments. It is likely that once such two shrinking protofilaments come side by side of a growing protofilament, this configuration can destabilize the middle one by the breaking lateral bonds, in this way making the lattice more unstable and finally forcing the entire microtubule  into a shrinking state. This prediction agrees with the conclusions of other authors \cite{odde,gardner} who have suggested, based on phenomenological arguments motivated by experimental data, that catastrophe is a multi-step process. In our view, the number of such steps precisely equals the number of protofilaments which need to become GDP-tipped in order for the microtubule catastrophe to be initiated. It is tempting to interpret this result in terms of the lateral bond energy o
 f protofilaments; however, it must be borne in mind that the model does not require that these GDP-tipped protofilaments need to be adjacent to each other. Clearly, this issue remains far from understood. Also, the implications of the age dependence of microtubule catastrophes, both { \it in vitro} and {\it in vivo} is another aspect of the problem which seems worthy of further investigation.

To summarize, the present model which treats a microtubule as 13 independent protofilaments, is fairly successful in predicting the time evolution and steady states of microtubule catastrophe frequency, for a range of growth rates. The highlight of this model is that it enables calculation of microtubule dynamic parameters starting from the dynamics of individual protofilaments, which, being a strictly one-dimensional problem, is more amenable to analytical treatments. This is not to argue that inter-protofilament interactions, neglected in the present model but included in several other studies, are unimportant; it is just that the limited experimental data available at the moment does not seem to be sufficient to make a clear-cut distinction between these two broad categories of models. The lateral bonds are also likely to play an important role in the process of rescue, which seems much less sensitive to tubulin concentration compared to catastrophe (see, e.g, \cite{howard
 review2012}), and much less understood from a modeling point of view. 

In {\it in vivo} situations, microtubules grow in a confined environment and typically encounter obstacles to growth in the form of rigid or flexible barriers. In the context of chromosome capture and subsequent spindle formation during the mitotic phase, it is well-known that microtubules exert forces both when polymerizing and depolymerizing. The negative effects of polymerization force, when generated against a rigid barrier, on the growth velocity of microtubules was first shown experimentally by Dogterom and Yurke \cite{dogterom-yurke} and subsequently studied theoretically by other authors(see, e.g., \cite{mogilner,vandoorn,fisher,kierfeld1}).  Experiments have also shown that the catastrophe frequency is enhanced by the proximity of the microtubule tip to a barrier, both {\it in vitro} \cite{janson} and {\it in vivo} \cite{komarova}, which is consistent with a reduced binding rate in the presence of force (see \cite{zhang,kierfeld2}, two recent theoretical studies of h
 ow microtubule dynamic instability is affected by force and confinement).  Interestingly, the 13-protofilament model used by the authors of \cite{mogilner,vandoorn,kierfeld1} is similar to the model studied in the second part of this paper. It should, therefore, be possible to extend the present model in a straightforward manner to include force-dependence of $k_g$, in a protofilament-specific manner \cite{mogilner,kierfeld1}.

\begin{acknowledgments}
The authors would like to acknowledge many fruitful discussions with the members of the Complex Fluids and Biological Physics group, Department of Physics, IIT Madras. We also acknowledge useful correspondence with D.N. Drechsel and the authors of Ref. \cite{ranjith} regarding the experimental data in \cite{drechsel}. 
\end{acknowledgments}

 \vspace{0.5cm}
\appendix*

\section{Expressions for $P_0$ and $P_1$ to ${\cal O}(r)$}

{\it Zero'th order terms:}

Since we perform a perturbative expansion of probabilities in $r$, and also $r$ is very small it is possible to retain terms up to first order in $r$.
In the calculations, we determine $P_0^{(0)}$ and $P_0^{(1)}$ independently, while the other required probabilities, especially  $P_1^{(0)}$ and $P_1^{(1)}$ in steady state are determined using the former. The dynamical equation for  $\phi^{(0)}(z,t)=\sum_{n=0}^{\infty} z^n P_n^{(0)}(t)$ is given by  
\begin{eqnarray}
\frac{\partial \phi^{(0)}(z,t)}{\partial t}=[k_h(\frac{1-z}{z})+kg (z-1)]\phi^{(0)}(z,t)\nonumber \\ -[k_h(\frac{1-z}{z})+(\nu_r^{\prime}-k_g)(1-z)]P_0^{(0)}(t). 
\label{eq:eqA1}
\end{eqnarray}

On solving equation Eq.\ref{eq:eqA1} using laplace transform technique with the initial condition that $P_N^{(0)}(t=0)=1$
we get, 
\begin{equation}
 \tilde \phi^{(0)}(z,s)= \frac {[z^N-\frac{1}{z}[k_h+(\nu_r^{\prime}-k_g )z](1-z)\tilde P_0^{(0)}(s)]}{[s-k_g(z-1)-k_h(\frac{(1-z)}{z})]} 
\label{eq:eqA2}
\end{equation}
 
\begin{equation}
\tilde \phi^{(0)}(z,s)= \frac {[k_h+(\nu_r^{\prime}-k_g)z](1-z)\tilde P_0^{(0)}(s)-z^{N+1}}{k_g(z-z_1)(z-z_2)},
\label{eq:eqA3}
\end{equation}
where the constants $z_1$ and $z_2$ are given by

\begin{equation}
z_{1,2}=\frac{(s+k_g+k_h)\mp \sqrt{(s+k_g+k_h)^2-4k_gk_h}} {2k_g}.
\label{eq:eq12}
\end{equation}

It is easily seen that $z_1<1$ while $z_2>1$, independent of the specific values of the parameters. Substituting  Eq.\ref{eq:eqA3}  in to Eq.\ref{eq:eq8} where $k=0$ in this case, we get,

\begin{widetext}
\begin{equation}
\tilde P_n^{(0)}(s)=\frac{1}{2\pi i}\oint[\frac{1}{(z_1-z)}-\frac{1}{(z_2-z)}]\frac{(k_h+(\nu_r^{\prime}-k_g)z)(1-z)\tilde P_0^{(0)}(s)-z^{N+1}}{k_g(z_2-z_1)z^{n+1}}dz.
\label{eq:eqA4}
\end{equation}
\end{widetext}

For $z<<z_1,z_2$, using binomial expansion we get the non vanishing terms as

\begin{widetext}
\begin{eqnarray}
\tilde P_n^{(0)}(s)=\frac{1}{k_g(z_2-z_1)}[z_1^{-n-1}[(k_h +(\nu_r^{\prime}-k_g)z_1)(1-z_1)\tilde P_0^{(0)}(s) -z_1^{N+1}]\nonumber \\ 
-z_2^{-n-1}[(k_h +(\nu_r^{\prime}-k_g)z_2)(1-z_2)\tilde P_0^{(0)}(s)-
z_2^{N+1}]].~~~~~~~~
\label{Eq:eqA5}
\end{eqnarray}
\end{widetext}

In the limit of $n\rightarrow\infty$, the expression for $\tilde P_n^{(0)}(s)$ should converge which is possible only if the coefficient of $z_1^{-n}$ vanishes, 
which fixes $\tilde P_0^{(0)}(s)$:

\begin{equation}
 \tilde P_0^{(0)}(s)= \frac {z_1^{N+1}}{(k_h+(\nu_r^{\prime}-k_g)z_1)(1-z_1)}.
 \label{eq:eqA6}
\end{equation}

{\it First order terms}:

The dynamical equation for  $\phi^{(1)}(z,t)=\sum_{n=0}^{\infty} z^n P_n^{(1)}(t)$ is given by
\begin{widetext}
 \begin{eqnarray}
 \frac{\partial \phi^{(1)}(z,t)}{\partial t}\nonumber =[k_h(\frac{1-z}{z})+kg (z-1)]\phi^{(1)}(z,t)-\frac{1}{z}[k_h+ (\nu_r^{\prime}-k_g )z](1-z)P_0^{(1)}(t)\\ 
-z \frac{\partial \phi^{(0)}(z,t)}{\partial z} +\sum_{i=0}^{\infty}z^i\sum_{m=i+1}^{\infty} P_{m}^{(0)}(t). ~~~~~~~~~~~~~~~~~\nonumber \\
\label{Eq:eqA9}
\end{eqnarray}
\end{widetext}
whose solution in Laplace space is 

\begin{widetext}
\begin{eqnarray}
\tilde \phi^{(1)}(z,s)= \frac {(k_h+(\nu_r^{\prime}-k_g)z)(1-z)\tilde P_0^{(1)}(s) +z^2\frac{\partial \tilde \phi^{(0)}(z,s)}{\partial z}
-\sum_{i=0}^{\infty}z^{i+1}\sum_{m=i+1}^{\infty}\tilde P_{m}^{(0)}(s)} {k_g(z-z_1)(z-z_2)},
\label{eq:eqA10}
\end{eqnarray} 
\end{widetext}
where, after explicit calculations, it is found that 

\begin{eqnarray}
 \sum_{i=0}^{\infty}z_1^i\sum_{m=i+1}^{\infty}\tilde P_{m}^{(0)}(s)=\frac{1}{k_g(z_2-z_1)}\bigg[ \frac{(1-z_1^N)}{(1-z_1)^2}\nonumber \\ -\frac{Nz_1^N}{(1-z_1)}+\frac{z_1^{N+1}(k_h+(\nu_r^{\prime}-k_g)z_2)}{(k_h+(\nu_r^{\prime}-k_g)z_1)(1-z_1)(z_2-z_1)} \nonumber \\ +\frac{1}{(z_2-1)(1-z_1)}-\frac{z_1^{N+1}}{(z_2-z_1)(1-z_1)}\bigg]\nonumber \\
 \label{eq:eqA7}
\end{eqnarray}

and

\begin{eqnarray}
 \frac{\partial \tilde \phi^{(0)}(z,s)}{\partial z}|_{z=z_1}=\frac{z_1^{N+1}}{k_g(z_1-z_2)^2(1-z_1)}+\frac{(N+1)z_1^{N}}{k_g(z_1-z_2)^2}\nonumber \\-\frac{N(N+1)z_1^{N-1}}{2k_g(z_1-z_2)}-\frac{(\nu_r^{\prime}-k_g)z_1^{N+1}}{k_g(z_1-z_2)^2(k_h+(\nu_r^{\prime}-k_g)z_1)}\nonumber \\ -\frac{(\nu_r^{\prime}-k_g)z_1^{N+1}}{k_g(z_1-z_2)(1-z_1)(k_h+(\nu_r^{\prime}-k_g)z_1)}.\nonumber \\
\label{eq:eqA8}
\end{eqnarray}

We now derive our required results. The general expression for $\tilde P_0^{(0)}(s)$ and  $\tilde P_0^{(1)}(s)$ are given by 
   
\begin{equation}
 \tilde P_0^{(0)}(s)= \frac {z_1^{N+1}}{(k_h+(\nu_r^{\prime}-k_g)z_1)(1-z_1)}.
 \label{eq:eqB1}
\end{equation}

\begin{widetext}
\begin{eqnarray}
\tilde P_0^{(1)}(s)= \bigg[ \frac{k_g}{(\nu_r^{\prime}-k_g+z_2k_g)(1-z_1)}\bigg]\bigg[\frac{(1-z_1^N)}{k_g^2(z_2-z_1)(1-z_1)^2}\nonumber \\ -\frac{Nz_1^N}{k_g^2(z_2-z_1)(1-z_1)}+\frac{z_1^{N+1}(k_h+(\nu_r^{\prime}-k_g)z_2)}{k_g^2(k_h+(\nu_r^{\prime}-k_g)z_1)(1-z_1)(z_2-z_1)^2}+ \frac{1}{k_g^2(z_2-z_1)(z_2-1)(1-z_1)}-\nonumber \\ -\frac{z_1^{N+1}}{k_g^2(z_2-z_1)^2(1-z_1)}-\frac{z_1^{N+2}}{k_g^2(z_1-z_2)^2(1-z_1)}-\frac{(N+1)z_1^{N+1}}{k_g^2(z_1-z_2)^2}+\frac{N(N+1)z_1^N}{2k_g^2(z_1-z_2)}+\nonumber \\ \frac{(\nu_r^{\prime}-k_g)z_1^{N+2}}{k_g^2(z_1-z_2)^2(k_h+(\nu_r^{\prime}-k_g)z_1)}+\frac{(\nu_r^{\prime}-k_g)z_1^{N+2}}{k_g^2(z_1-z_2)(1-z_1)(k_h+(\nu_r^{\prime}-k_g)z_1)}\bigg].\nonumber \\
\label{eq:eqB2}
\end{eqnarray}
\end{widetext}

The steady state expressions for  $ P_1^{(0)}$ and  $ P_1^{(1)}$ can be determined using the expressions given below, which follow directly from the master equation:
\begin{eqnarray}
   P_1^{(0)}=\frac{\nu_r^{\prime}}{k_h } P_0^{(0)}~~~~~~~~~~~~~~~~~~~~~\\ \nonumber
 P_1^{(1)}=\frac{1}{k_h}\bigg[\nu_r^{\prime}  P_0^{(1)}+ P_0^{(0)}-1\bigg].
\end{eqnarray}

\begin{thebibliography}{100}

\bibitem{desai}A. Desai and T. Mitchison, Annu. Rev. Cell Dev. Biol. {\bf13}, 83 (1997).

\bibitem{mitchison}T. Mitchison and M. Kirschner, Nature  {\bf 312}, 232 (1984).

\bibitem{howardreview2012} M. Gardner, M. Zanic and J. Howard, Curr. Opin. Cell  Biol. {\bf 25}, 1 (2012).

\bibitem{caplow} Caplow M. and Shanks,  Mol. Biol. Cell {\bf 7}, 663 (1996).
  
\bibitem{schek} H. T. Schek III, M. K. Gardner, J. Cheng, D. J. Odde and A. J. Hunt, Curr. Biol. {\bf17}, 1445 (2007).

\bibitem{hill} T. L. Hill and Y. Chen, Proc. Natl. Acad. Sci. USA {\bf81}, 5772 (1984).

\bibitem{FHL} H. Flyvbjerg, T. E. Holy, and S. Leibler, Phys. Rev. Lett.  {\bf73}, 2372 (1994); Phys. Rev. E {\bf 54}, 5538 (1996).

\bibitem{margolin} G. Margolin, I. V. Gregoretti, H. V. Goodson, and M. S. Alber, Phys. Rev. E {\bf74}, 041920 (2006).

\bibitem{antal} T. Antal, P. L. Krapivsky, S. Redner, M. Mailman, and B. Chakraborty, Phys. Rev. E {\bf76}, 041907 (2007).

\bibitem{ranjith}R. Padinhateeri, A. B. Kolomeisky, and D. Lacoste, Biophys. J. {\bf102}, 1274 (2012).

\bibitem{vanburen} V. VanBuren, D. J. Odde, and L. Cassimeris,  Proc. Natl. Acad. Sci. USA {\bf99}, 6035 (2002); V. VanBuren, L. Cassimeris, and D. J. Odde, Biophys. J. {\bf89}, 2911 (2005).

\bibitem{molodtsov} M. I. Molodtsov, E. L. Grishchuk, A. K. Efremov, J. R. McIntosh, and F. I. Ataullakhanov, Proc. Natl. Acad. Sci. USA {\bf 102}, 4353 (2005).

\bibitem{piette} B. M. Piette et al., PLoS ONE 4, e6378 (2009).

\bibitem{nedelec} L. Brun, B. Rupp, J. J. Ward, and F. N\'{e}d\'{e}lec, Proc. Natl. Acad. Sci. USA {\bf106}, 21173 (2009).

\bibitem{alber} G. Margolin et al., Mol. Biol. Cell {\bf 23}, 642 (2012).

\bibitem{Li} X. Li, R. Lipowsky, and J. Kierfeld, Europhys. Lett. {\bf 89}, 38010 (2010).

\bibitem{ranjith2010} P. Ranjith, K. Mallick, J -F. Joanny, and D. Lacoste, Biophys. J. {\bf 98}, 1418 (2010).

\bibitem{supp} See Supplemental Material at [URL will be inserted by publisher] for details of the calculations.

\bibitem{Gill}  D. T. Gillespie, J. Phys. Chem. {\bf 81}, 2340 (1977).

\bibitem{math} Wolfram Research, Inc., Mathematica, Version 7.0, Champaign, IL (2008).

\bibitem{foot} In principle, rescue could also occur due to a $T$-island inside the filament getting `exposed' after dissociation of all the preceding $D$-mers; in our work, we disregard this possibility like most other authors, see, however, the discussions in \cite{howardreview2012}.
   
 \bibitem{drechsel} D. N. Drechsel, A. A. Hyman, M. H. Cobb, and M. W. Kirschner, Mol. Biol. Cell {\bf3}, 1141 (1992).

\bibitem{anderson} H. Bowne-Anderson, M. Zanic, M. Kauer, and J. Howard, Bioessays {\bf 35}, 452 (2013).
 
 \bibitem{odde} D. J. Odde, L. Cassimeris, and H. M. Buettner, Biophys. J. {\bf 69}, 796 (1995).
 
 \bibitem{stepanova} T. Stepanova et al., Curr. Biol. {\bf 20}, 1023 (2010).

\bibitem{gardner} M. K. Gardner, M. Zanic, C. Gell, V. Bormuth, and J. Howard, Cell {\bf147}, 1092  (2011).

\bibitem{walker} R. A. Walker, E. T. \"{O}Brien, N. K. Pryer, M. F. Soboeiro, W. A. Voter, H. P. Erickson, and E. D. Salmon, J. Cell Biol. {\bf 107}, 1437 (1988).

\bibitem{janson} M. E. Janson, M. E. de Dood and M. Dogterom, J. Cell Biol. {\bf 161}, 1029 (2003).

\bibitem{dogterom-yurke} M. Dogterom and B. Yurke, Science {\bf 278}, 856 (1997).

\bibitem{mogilner} A. Mogilner and G. Oster, Eur. Biophys. J. {\bf 28}, 235 (1999).

\bibitem{vandoorn} G. S. van Doorn, C. Tanase, B. M. Mulder, and M. Dogterom, Eur. Biophys. J. {\bf 29}, 2 (2000).

\bibitem{fisher} A. B. Kolomeisky and M. E. Fisher, Biophys. J. {\bf 80}, 149 (2001). 

\bibitem{kierfeld1} J. Krawczyk and J. Kierfeld, Europhys. Lett. {\bf 93}, 28006 (2011).

\bibitem{komarova} D. R. Drummond and R.A. Cross, Curr. Biol. {\bf 10}, 766 (2000); A. Komarova, I. A. Vorobjev, and G. G. Borisy, J. Cell Sci. {\bf 115}, 3527 (2002); D. Foethke, T. Makushok, D. Brunner, and F. N\'{e}d\'{e}lec, Mol. Syst. Biol. {\bf 5}, 241 (2009); C. Tischer, D. Brunner, and M. Dogterom, Mol. Syst. Biol. {\bf 5}, 250 (2009).

\bibitem{zhang} Y. Zhang, J. Biol. Chem. {\bf 286}, 39439 (2011).

\bibitem{kierfeld2} B. Zelinski, N. M\"{u}ller, and J. Kierfeld, Phys. Rev. E {\bf 86}, 041918 (2012).

 \end{thebibliography}
\end{document}